\documentclass[prl,superscriptaddress,reprint,showpacs]{revtex4-1}
\usepackage{graphics}
\newcommand{\ket}[1]{\ensuremath{\left|#1\right>}}%
\newcommand{\figref}[2]{Fig. \ref{#1}{\bf{}#2}}%
\newcommand{\tmp}{\ensuremath{\mathcal{T}}}%

\begin{document}
\title{Electrometry using coherent exchange oscillations in a
  singlet-triplet qubit}
\date{\today}
\author{O. E. Dial}
\author{M. D. Shulman}
\author{S. P. Harvey}
\affiliation{Department of Physics, Harvard University, Cambridge, MA, 02138, USA}
\author{H. Bluhm}
\affiliation{Department of Physics, Harvard University, Cambridge, MA, 02138, USA}
\affiliation{Current Address: 2nd Institute of Physics C, RWTH Aachen University, 52074 Aachen, Germany}
\author{V. Umansky}
\affiliation{Braun Center for Submicron Research, Department
of Condensed Matter Physics, Weizmann Institude of Science, Rehovot
76100 Israel}
\author{A. Yacoby}
\affiliation{Department of Physics, Harvard University, Cambridge, MA, 02138, USA}
\pacs{85.35.Be 03.67.-a 07.50.Ls 76.60.Lz 07.50.Hp}
\begin{abstract}
Two level systems that can be reliably controlled and measured
hold promise as qubits both for metrology
 and for quantum information
science (QIS). Since a fluctuating environment
limits the performance of qubits in both capacities, understanding the
environmental coupling and dynamics is key to improving qubit
performance. We show measurements of the level splitting and dephasing
due to voltage noise of a GaAs singlet-triplet qubit 
during exchange oscillations. Unexpectedly, the voltage fluctuations
are non-Markovian even at high frequencies and exhibit a strong
temperature dependence. The magnitude of the fluctuations allows the
qubit to be used as a charge sensor with a sensitivity of $2 \times
10^{-8} e/\sqrt{\mathrm{Hz}}$, two orders of magnitude better than a
quantum-limited RF single electron transistor (RF-SET). Based on these measurements we provide
recommendations for improving qubit coherence, allowing for higher
fidelity operations and improved charge sensitivity.
\end{abstract}
\maketitle

\begin{figure}
\resizebox{\columnwidth}{!}{\includegraphics{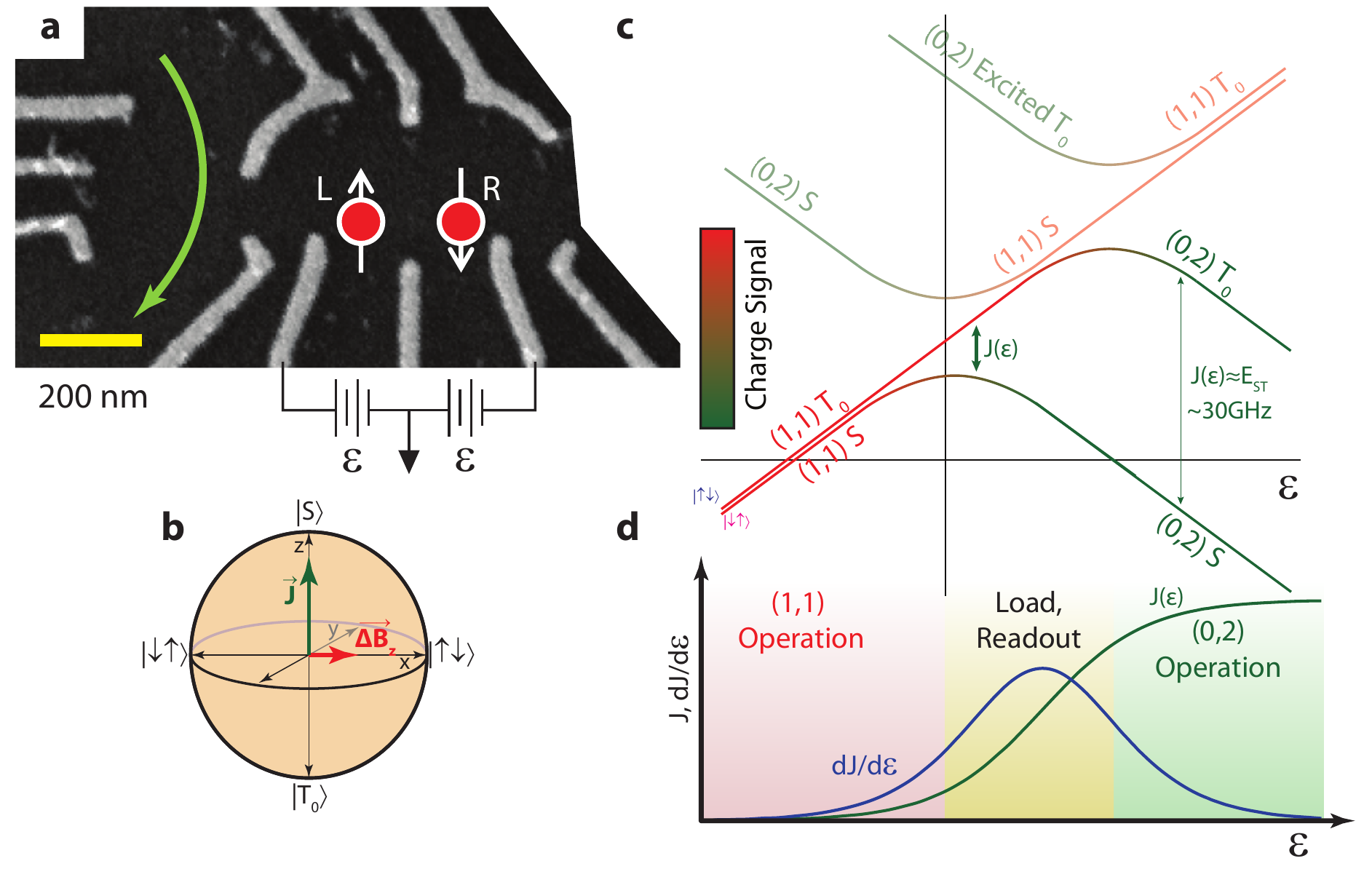}}
\caption{ The device used in these measurements is a
  gate-defined $S$-$T_0$ qubit with an integrated RF sensing dot.  \textbf{a} The
  detuning $\epsilon$ is the voltage applied to the dedicated
  high-frequency control leads pictured.  \textbf{b}, The Bloch sphere
  that describes the logical subspace of this device features two
  rotation axes ($J$ and $\Delta{}B_{Z}$) both controlled with DC
  voltage pulses. \textbf{c}, An energy diagram of the relevant low-lying
  states as a function of $\epsilon$.  States outside of the logical
  subspace of the qubit are grayed out.  \textbf{d}, $J(\epsilon)$ and
  $dJ/d\epsilon$ in three regions; the $(1,1)$ region where $J$ and
  $dJ/d\epsilon$ are both small and $S$-$T_0$ qubits are typically
  operated, the transitional region where $J$ and $dJ/d\epsilon$ are
  both large where the qubit is loaded and measured, and the $(0,2)$
  region where $J$ is large but $dJ/d\epsilon$ is small and large
  quality oscillations are possible.
\label{pulses}
}
\end{figure}

Two level quantum systems (qubits) are emerging as promising
candidates both for quantum information processing
\cite{DiVincenzo2000} and for sensitive metrology
\cite{Chernobrod2005,Maletinsky2012}.  When prepared in a
superposition of two states and allowed to evolve, the state of the
system precesses with a frequency proportional to the splitting
between the states. However, on a timescale of the coherence time,
$T_2$, the qubit loses its quantum information due to interactions
with its noisy environment. This causes qubit oscillations to decay
and limits the fidelity of quantum control and the precision of
qubit-based measurements.  In this work we study singlet-triplet
($S$-$T_0$) qubits, a particular realization of spin qubits
\cite{Hanson2007, Loss1998,Koppens2006,Pioro2008,Nowack2007,Pioro2007,Churchill2009,Nadj-Perge2010},
which store quantum information in the joint spin state of two
electrons\cite{Levy2002,Petta2005,Taylor2007}. We form the qubit in
two gate-defined lateral quantum dots (QD) in a GaAs/AlGaAs
heterostructure (\figref{pulses}{a}). The QDs are depleted until there
is exactly one electron left in each, so that the system occupies the
so-called $(1,1)$ charge configuration. Here $(n_L,n_R)$ describes a
double QD with $n_L$ electrons in the left dot and $n_R$ electrons in
the right dot.  This two-electron system has four possible spin
states: \ket{S}, \ket{T_{+}}, \ket{T_{0}}, and \ket{T_{-}}.  The
{\ket{S},\ket{T_0}} subspace is used as the logical subspace for this
qubit because it is insensitive to homogeneous magnetic field
fluctuations and is manipulable using only pulsed DC electric fields
\cite{Petta2005,Taylor2005,Levy2002}. The relevant low-lying energy
levels of this qubit are shown in \figref{pulses}{c}.  Two distinct
rotations are possible in these devices: rotations around the $x$-axis
of the Bloch sphere driven by difference in magnetic field between the
QDs, $\Delta{}B_z$(provided in this experiment by feedback-stabilized
hyperfine interactions\cite{Bluhm2009}), and rotations around the
$z$-axis driven by the exchange interaction, $J$ (\figref{pulses}{b})
\cite{Foletti2009}.  A \ket{S} can be prepared quickly with high
fidelity by exchanging an electron with the QD leads, and the
projection of the state of the qubit along the $z$-axis can be
measured using RF reflectometery with an adjacent sensing QD (green
arrow in \figref{pulses}{a})\cite{Reilly2007,Barthel2009}.

Previous work on $S$-$T_{0}$ qubits focused almost entirely on $x$
($\Delta{}B_{Z}$) rotations, which are dephased by fluctuations in the
nuclear bath \cite{Bluhm2011,Barthel2010b,Medford2012}. In this work,
we focus on the exchange interaction, which creates a splitting, $J$,
between the \ket{S} and \ket{T_0} states once the $(1,1)$ and
$(0,2)$\ket{S} states of the double QD are brought near resonance
(\figref{pulses}{c}). The value of $J$ depends on the energy detuning,
$\epsilon$, between the QDs. The exchange interaction drives single
qubit rotations in $S$-$T_{0}$ \cite{Petta2005} and exchange-only
\cite{Laird2010,Gaudreau2011} qubits and is the foundation for
two-qubit operations in single spin, $S$-$T_0$, and exchange-only
qubits
(\cite{Shulman2012,VanWeperen2011,Loss1998,Brunner2011,Nowack2011}).
Exchange oscillations are dephased by fluctuations in $J$
(\figref{pulses}{c}) driven, for example, by $\epsilon$ (voltage)
fluctuations between the dots with a tunable sensitivity proportional
to $dJ/d\epsilon$ (\figref{pulses}{d}) \cite{Culcer2009}.  We will
show that this controllable sensitivity is a useful experimental tool
for probing the noise bath dynamics.  Previous studies have shown the
decay of exchange oscillations within a few $\pi$ rotations
\cite{Petta2005,Maune2011}, but a detailed study of the nature of the
noise bath giving rise to this decay is still lacking.  In this work,
using nuclear feedback to control $x$-rotations, we systematically
explore the low frequency noise portion of the voltage noise bath and
its temperature dependence, as well as introduce a new Hahn-echo based
measurement of the high frequency components of the voltage noise and
its temperature dependence.

\begin{figure*}
\resizebox{\textwidth}{!}{\includegraphics{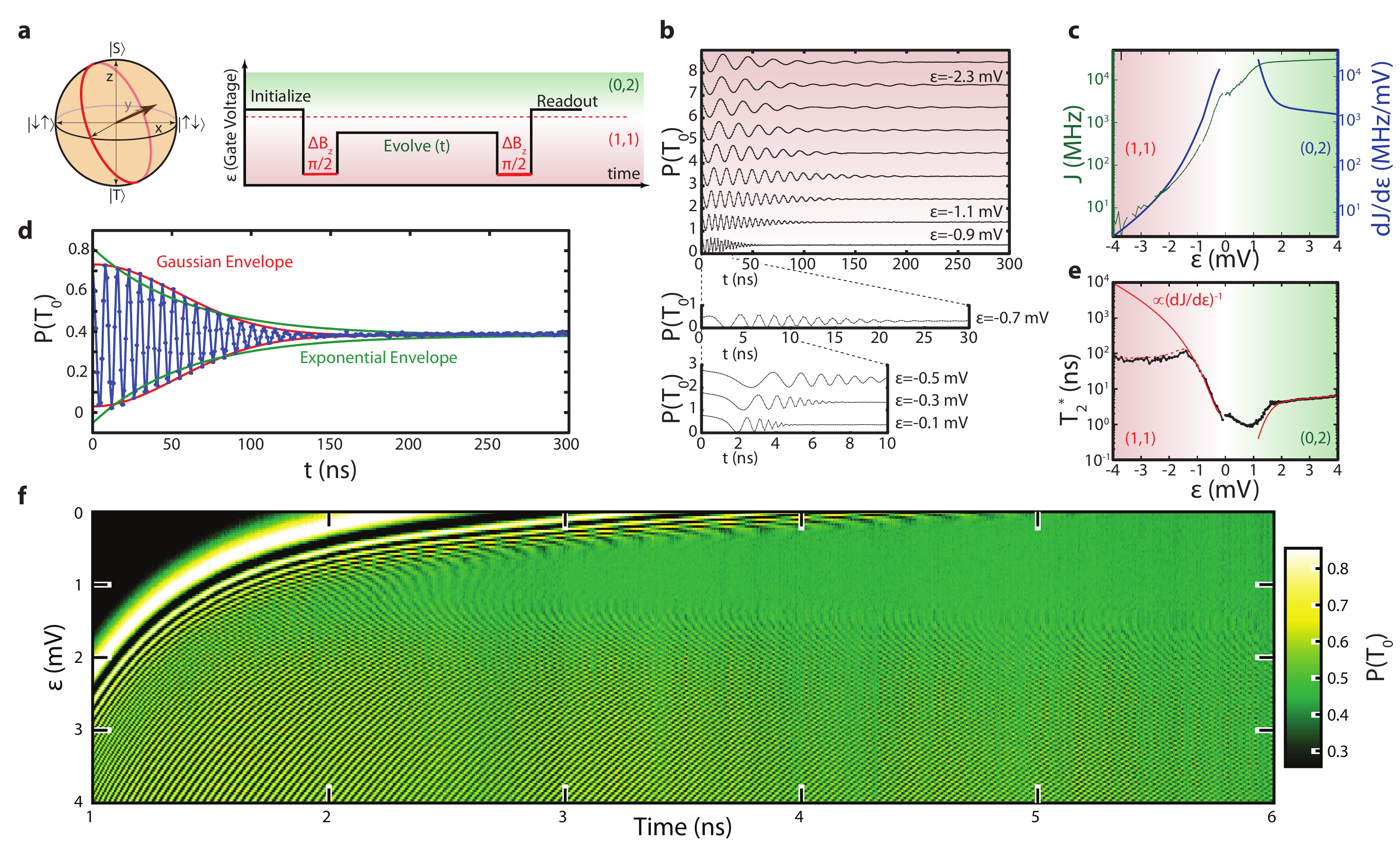}}
\caption{Ramsey oscilllations reveal low frequency enivronmental dynamics. \textbf{a}, The pulse sequence used to measure exchange
  oscillations uses a stabilized nuclear gradient to prepare and
  readout the qubit and gives good contrast over a wide range of
  $J$. \textbf{b}, Exchange oscillations measured over a variety of
  detunings $\epsilon$ and timescales consistently show larger $T_2^*$
  as $dJ/d\epsilon$ shrinks until dephasing due to nuclear
  fluctuations sets in at very negative $\epsilon$. \textbf{c},
  Extracted values of $J$ and $dJ/d\epsilon$ as a function of
  $\epsilon$. \textbf{d}, The decay curve of FID exchange oscillations
  shows Gaussian decay. \textbf{e}, Extracted values of $T_{2}^{*}$
  and $dJ/d\epsilon$ as a function of $\epsilon$.  $T_{2}^{*}$ is
  proportional to $(dJ/d\epsilon)^{-1}$, indicating that voltage noise
  is the cause of dephasing of charge oscillations. \textbf{f}, Charge
  oscillations measured in $(0,2)$. This figure portrays the three basic
  regions we can operate our device in: a region of low frequency
  oscillations and small $dJ/d\epsilon$, a region of large frequency
  oscillations and large $dJ/d\epsilon$, and a region where
  oscillations are fast but $dJ/d\epsilon$ is comparatively small.
\label{t2star}
}
\end{figure*}

The simplest probe of $J$ and its fluctuations is a free induction
decay (FID) experiment, in which the qubit is allowed to freely
precess for a time $t$ under the influence of the exchange
splitting. For FID measurements, we use a $\pi/2$ pulse around the
$x$-axis to prepare and readout the state of the qubit along the
$y$-axis (\figref{t2star}{a}, Fig. \textbf{S1}). \figref{t2star}{b}
shows qubit oscillations as a function of $t$ for many different
values of $\epsilon$. By measuring the period of these oscillations
with $t$ we extract $J(\epsilon)$, and we calculate $dJ/d\epsilon$ by
fitting $J(\epsilon)$ to a smooth function and differentiating
(\figref{t2star}{c}). For negative $\epsilon$ (small $J$), we
empirically find across many devices and tunings that $J$ is well
described by $J(\epsilon) \simeq J_0 + J_1 exp(-\epsilon/\epsilon_0)$.

The oscillations in these FID experiments decay due to voltage noise
from DC up to a frequency of approximately $1/t$.  As the relaxation
time, $T_1$ is in excess of 100$\mu$s in this regime, $T_1$ decay is
not an important source of decoherence (Fig. \textbf{S4}).  The shape
of the decay envelope and the scaling of coherence time with
$dJ/d\epsilon$ (which effectively changes the magnitude of the noise)
reveal information about the underlying noise spectrum. White
(Markovian) noise, for example, results in an exponential decay of
$e^{-t/T_2^{*}}$ where $T_2^*\propto (dJ/d\epsilon)^{-2}$ is the
inhomogeneously broadened coherence time \cite{Cywinski2008B}.
However, we find that the decay is Gaussian (\figref{t2star}{d}) and
that $T_2^*$ (black line in \figref{t2star}{e}) is proportional to
$(dJ/d\epsilon)^{-1}$ (red solid line in \figref{t2star}{e}) across
two orders of magnitude of $T_2^*$. Both of these findings can be
explained by quasistatic noise, which is low frequency compared to
$1/T_2^*$. In such a case, one expects an amplitude decay of the form
$\exp{\left[-(t/T_2^*)^2\right]}$, where
$T_{2}^{*}=\frac{1}{\sqrt{2} \pi (dJ/d\epsilon) \epsilon_{RMS}}$ and
$\epsilon_{RMS}$ is the root-mean-squared fluctuation in $\epsilon$
(Eq. \textbf{S3}). From the ratio of $T_2^*$ to $(dJ/d\epsilon)^{-1}$,
we calculate $\epsilon_{RMS}$ $=$ 8$\mu$V in our device.  At very
negative $\epsilon$, $J$ becomes smaller than $\Delta B_z$, and
nuclear noise limits $T_2^*$ to approximately 90ns, which is
consistent with previous work \cite{Bluhm2009}. We confirm that this
effect explains deviations of $T_2^*$ from $(dJ/d\epsilon)^{-1}$ by
using a model that includes the independently measured
$T_{2,nuclear}^{*}$ and $\Delta B_z$ (Eq. \textbf{S1}) and observe
that it agrees well with measured $T_{2}^{*}$ at large negative
$\epsilon$ (dashed red line in \figref{t2star}{e}).

Since we observe $J$ to be approximately an exponential function of
$\epsilon$, ($dJ/d\epsilon \sim J$), we expect and observe the quality
(number of coherent oscillations) of these FID oscillations, $Q\equiv
J T_{2}^{*}/2\pi \sim J(dJ/d\epsilon )^{-1} $, to be approximately
constant regardless of $\epsilon$. However, when $\epsilon$ is made
very positive and $J$ is large, an avoided crossing occurs between the
$(1,1)$\ket{T_0} and the $(0,2)$\ket{T_0} state, making the $(0,2)$\ket{S}
and $(0,2)$\ket{T_0} states electrostatically virtually identical. Here,
as $\epsilon$ is increased, $J$ increases but $dJ/d\epsilon$
decreases(\figref{pulses}{d}), allowing us to probe high quality
exchange rotations and test our charge noise model in a regime that
has never before been explored.

Using a modified pulse sequence that changes the clock frequency of
our waveform generators to achieve picosecond timing resolution
(Fig. \textbf{S1})), we measure exchange oscillations in $(0,2)$ as a
function of $\epsilon$ and time (\figref{t2star}{e}) and we extract
both $J$ (\figref{t2star}{c}) and $T_{2}^{*}$ (\figref{t2star}{d}) as
a function of $\epsilon$. Indeed, the predicted behavior is observed:
for moderate $\epsilon$ we see fast oscillations that decay after a
few ns, and for the largest $\epsilon$ we see even faster oscillations
that decay slowly. Here, too, we observe that $T_{2}^{*} \propto
(\frac{dJ}{d\epsilon})^{-1}$ (\figref{t2star}{d}), which indicates
that FID oscillations in $(0,2)$ are also primarily dephased by low
frequency voltage noise. We note, however, that we extract a different
constant of proportionality between $T_{2}^{*}$ and
$(dJ/d\epsilon)^{-1}$ for $(1,1)$ and $(0,2)$. This is expected given that
the charge distributions associated with the qubit states are very
different in these two regimes and thus have different sensitivities to
applied electric fields.  We note that in the regions of largest
$dJ/d\epsilon$ (near $\epsilon=0$), $T_2^{*}$ is shorter than the rise
time of our signal generator and we systematically underestimate $J$
and overestimate $T_2^{*}$ (Fig. \textbf{S1}).

The above measurements indicate that the dephasing during FID
experiments in both $(1,1)$ and $(0,2)$ arises overwhelmingly due to low
frequency (non-Markovian) noise, and the observed linear dependence of
$T_2^*$ on $(dJ/d\epsilon)^{-1}$ strongly suggests that $\epsilon$
noise is indeed responsible for the observed dephasing, as these data
rule out dephasing from other mechanisms in most realistic situations
(see supplement sec. 5). In the presence of such low frequency noise,
the addition of a $\pi$-pulse half-way through the free evolution can
partially decouple the qubit from its noisy environment.  Such a
``Hahn-echo'' \cite{Hahn1950} sequence prolongs coherence, which is
useful for complex quantum operations \cite{Shulman2012}, sensitive
detection \cite{Maze2008}, and probing higher-frequency portions of
the voltage noise bath.  Rather than being sensitive to noise from DC
to $1/\tau$ where $\tau$ is the total evolution time, these echo
sequences have a noise sensitivity peaked at $f \approx 1/\tau$ and a
reduced sensitivity at lower frequencies.

\begin{figure}
\resizebox{\columnwidth}{!}{\includegraphics{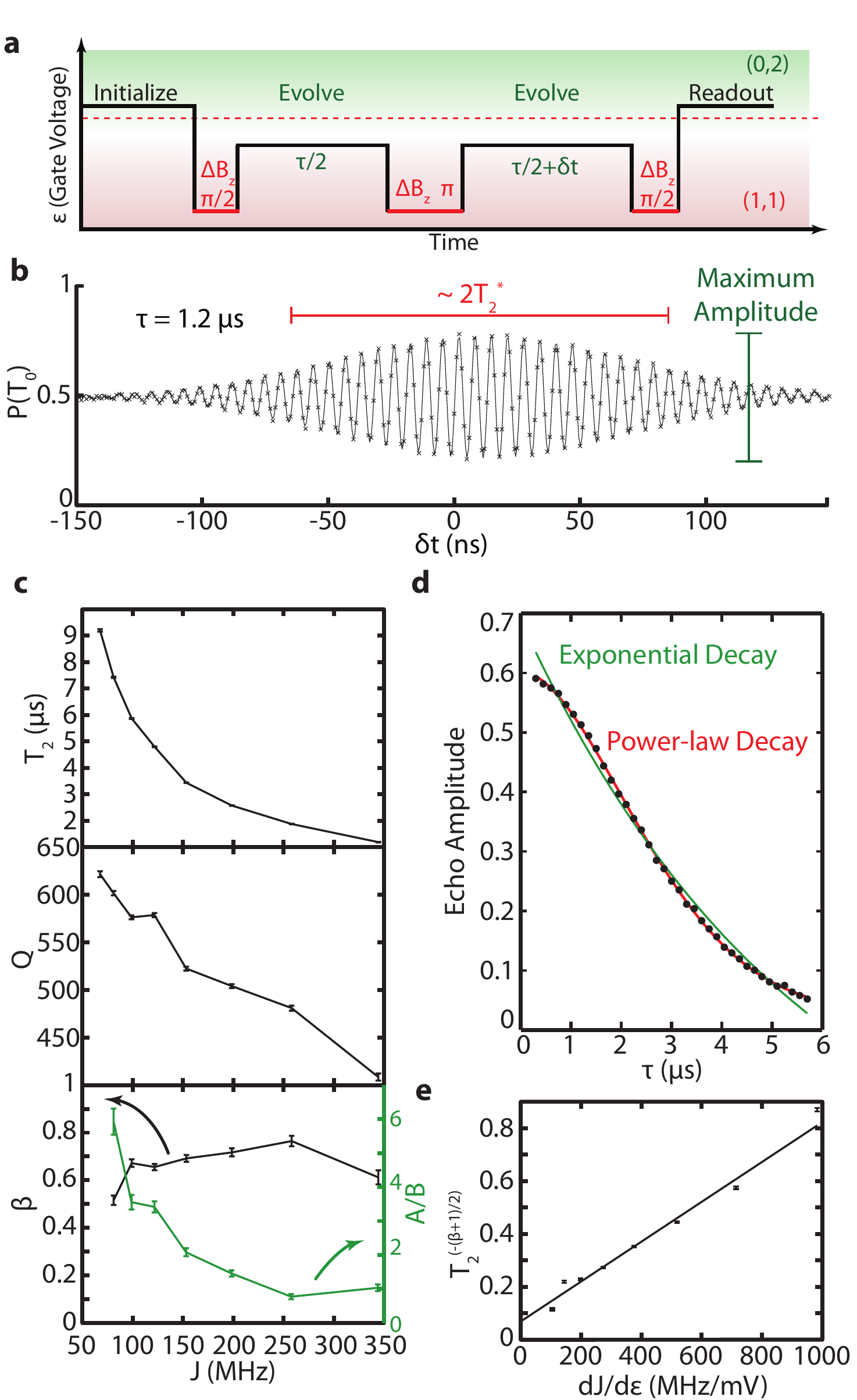}}
\caption{Spin-echo measurements reveal high frequency bath dynamics.  \textbf{a}, The pulse sequence used to measure exchange echo
  rotations. \textbf{b}, A typical echo signal.  The overall shape of
  the envelope reflects $T_2^*$, while the amplitude of the envelope
  as a function of $\tau$ (not pictured) reflects $T_2^{echo}$.
  \textbf{c}, $T_{2}^{echo}$ and $Q\equiv J T_{2}^{echo}/2\pi$ as a
  function of $J$. A comparison of the two noise models: power law and
  a mixture of white and $1/f$ noise. Noise with a power law spectrum
  fits over a wide range of frequencies (constant $\beta$), but the
  relative contributions of white and $1/f$ noise change as a function
  of $\epsilon$. \textbf{d}, A typical echo decay is non-exponential
  but is well fit by
  $\exp(-(\tau/T_{2}^{echo})^{\beta+1})$. \textbf{e}, $T_2^{echo}$
  varies with $dJ/d\epsilon$ in a fashion consistent with dephasing
  due to power law voltage fluctuations.
\label{echo}
}
\end{figure}

In our echo measurements, we select a fixed $\epsilon$ inside $(1,1)$
for the free evolutions, and we sweep the length of the evolution
following the $\pi$-pulse time by small increments $\delta t$ to
reveal an echo envelope (\figref{echo}{a-b}).  The maximum amplitude
of this observed envelope reveals the extent to which the state has
dephased during the echo process, while the Gaussian shape and width
of the envelope arise from an effective single-qubit rotation for a
time $\delta t$, and thus reflect the same $T_2^*$ and low frequency
noise measured in FID experiments.  We note that this exchange echo is
distinct from the echo measurements previously performed in
singlet-triplet qubits\cite{Bluhm2011,Barthel2010b,Medford2012} in
that we use $\Delta{}B_z$ rotations to echo away voltage noise, rather
than $J$ rotations to echo away noise in the nuclear bath.

The use of Hahn echo dramatically improves coherence times, with
$T_2^{echo}$ (the $\tau$ at which the observed echo amplitude has
decayed by $1/e$) as large as 9 $\mu{}s$, corresponding to qualities
($Q\equiv T_{2}^{echo} J/2\pi$) larger than 600 (\figref{echo}{c}).
If at high frequencies (50kHz-1MHz) the voltage noise were white
(Markovian), we would observe exponential decay of the echo amplitude
with $\tau$. However, we find that the decay of the echo signal is
non-exponential (\figref{echo}{d}), indicating that even in this
relatively high-frequency band being probed by this measurement, the
noise bath is not white.

A simple noise model that can account for this decay includes a
mixture of white and $1/f$ noise, $S_{\epsilon}(f) = A + B/f$, which
leads to an echo amplitude decay $exp(-\tau/C_0 - \tau^2/C_1)$
\cite{Cywinski2008B}, where $C_{0,1}$ are functions of the noise
power. Since $C_{0,1}$ are both proportional to $(dJ/d\epsilon)^{-2}$,
we expect the ratio $C_0/C_1 \propto A/B$ to be independent of
$dJ/d\epsilon$. While this decay accurately describes the decay for a
single value of $\epsilon$, as we change $\epsilon$ (and therefore
$dJ/d\epsilon$), we find that the ratio $A/B$ changes, indicating that
this model is inconsistent with our data because the relative
contributions of white and $1/f$ noise change
(\figref{echo}{c}). Alternatively, we consider a power law noise model
$S_{\epsilon}(f) = \frac{S_0}{f^\beta}$, which leads to an echo
amplitude decay $exp(-(\tau/T_2^{echo})^{\beta+1})$. With this model
we expect $\beta$ to be independent of $dJ/d\epsilon$, and we indeed
observe $\beta \approx 0.7$ for all values of $\epsilon$
(\figref{echo}{c}), indicating that this model can adequately describe
our observed noise from approximately 50 kHz to 1 MHz.  We further
confirm that the observed dephasing is consistent with voltage noise
by checking $T_{2}^{echo}$ has the expected dependence on
$dJ/d\epsilon$, namely, $T_2^{echo} \propto
(dJ/d\epsilon)^{\frac{-2}{\beta + 1}}$ (\figref{echo}{e}). From the
scale factor, we deduce that the noise is well approximated by
$S_\epsilon(f) = 8 \times 10^{-16} \frac{\mathrm{V}^2}{\mathrm{Hz}}
\left(\frac{\mathrm{1Hz}}{f}\right)^{0.7}$ from approximately 50 kHz
to 1 MHz, corresponding to $\epsilon$ noise of 0.2
nV/$\sqrt{\mathrm{Hz}}$ at 1 MHz. We note that this noise exceeds that
accounted for by known sources of noise present in the experiment,
including instrumental noise on the device gates and Johnson noise of
the wiring.  The RMS noise deduced from our FID measurements exceeds that
expected from this power-law noise; there is excess noise at very low frequencies
in the device.

\begin{figure}
\resizebox{\columnwidth}{!}{\includegraphics{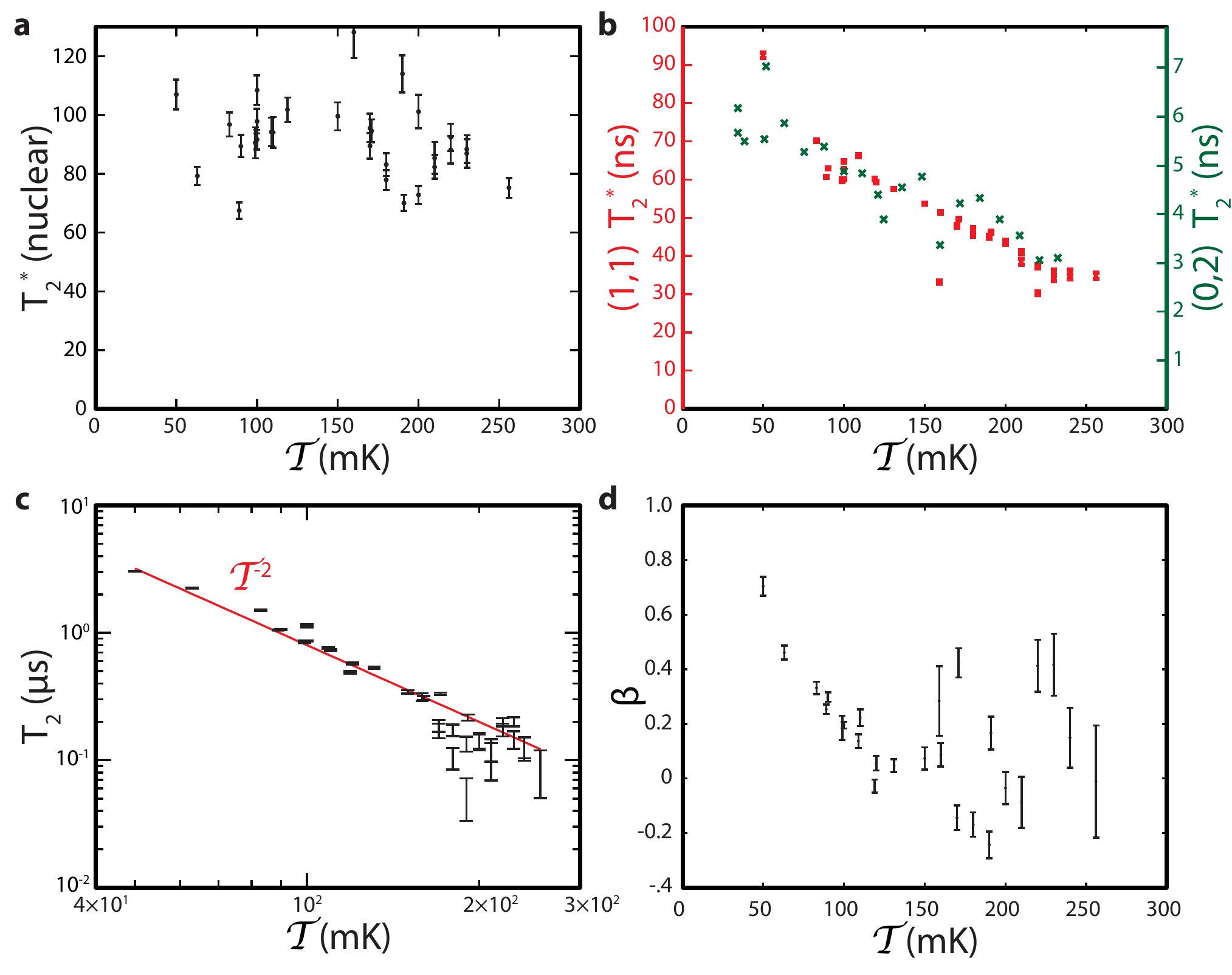}}
\caption{Temperature dependence indicates the noise leading to dephasing originates near or inside the device. \textbf{a}, $T^*_{2,nuclear}$ does not depend significantly
  on \tmp.  \textbf{b}, $T_2^*$ in $(1,1)$ (red squares) and $(0,2)$ (green
  crosses) has the same, weak, scaled dependence on \tmp.  \textbf{c},
  $T_2^{echo}$ shows a strong temperature dependence near $\tmp^2$
  (red line) across over an order of magnitude in coherence times.
  \textbf{e}, As \tmp is increased, $\beta$ approaches zero indicating
  the noise leading to decoherence becomes nearly white.  Uncertainties
  in $\beta$ are larger at higher temperatures due to the fast decay
  of the echo.
\label{temp}
}
\end{figure}

Thus far, we have explored the voltage noise bath at the base
temperature of our dilution refrigerator (\tmp$\approx$50 mK). We gain
additional insight into the properties of the voltage noise by
studying its temperature dependence. For the nuclear bath, attainable
temperatures in dilution refrigerators are much larger than the
nuclear Zeeman splitting and no temperature dependence is expected.
This is confirmed by measuring $T_{2,{nuclear}}^*$, the dephasing time
for FID rotations in the stabilized nuclear gradient field (at $J$=0)
as a function of \tmp (\figref{temp}{a}).

By contrast, $T_2^*$ in $(1,1)$ and $(0,2)$ show unexpected temperature
dependences (\figref{temp}{b}).  These have the same scaled
temperature dependence: $T_{2,(1,1)}^*(\tmp) \propto
T_{2,(0,2)}^*(\tmp)$, suggesting the loss of coherence is due to the
same mechanism, presumably increased voltage noise, in both instances.
In both cases $T_2^*$ is roughly linear with \tmp, indicating that
only small gains are likely to be made in the quality of FID-based
rotations by reducing \tmp. By comparison, $T_2^{echo}$ shows a strong
dependence of $T_2^{echo} \propto \tmp^{-2}$ (\figref{temp}{c}).  As
\tmp is increased, the observed noise becomes increasingly white
(frequency independent) (\figref{temp}{d}), though measurements of
$\beta$ become inaccurate at large \tmp where $T_{2}^{echo}$ is small.
We note that the underlying mechanism of this temperature dependence
is currently unknown, however the dependence of the observed dephasing
on temperature strongly suggests that the noise originates within the
device rather than the experimental apparatus.  Lower temperatures
carry a double benefit for echo coherence; the noise becomes both
smaller and more non-Markovian, thereby increasing coherence times and
extending the potential for multi-pulse dynamical decoupling sequences
to mitigate the effects of the noise \cite{Carr1954,Uhrig2007}.  This
trend shows no indication of saturating at low temperatures; it
appears likely much longer coherence times are attainable by reducing
temperatures with more effective refrigeration (Fig. \textbf{S3}).

Operating at base temperature and using the Hahn echo sequence
described above, we observe a voltage sensitivity of 0.2
nV/$\sqrt{\mathrm{Hz}}$ at 1 MHz, which suggests that the qubit can be
used as a sensitive electrometer at this frequency. In order to compare
to other electrometers, we convert our voltage sensitivity into a
charge sensitivity of $2 \times 10^{-8} \mathrm{e} /
\sqrt{\mathrm{Hz}}$ by dividing by the Coulomb blockade peak spacing,
10mV. This value is nearly two orders of magnitude better than the
theoretical limits for RF-SETs\cite{Devoret2000}, and is limited only
by $T_{2}^{echo}$, which the data suggest can be improved.

Using both FID and echo measurements, we have characterized the
exchange interaction and presented experimental evidence that exchange
rotations in $S$-$T_0$ qubits dephase due to voltage noise. These
measurements reveal that the voltage noise bath that couples to the
qubit is non-Markovian and establish baseline noise levels for
$S$-$T_0$ qubits. We suggest that further improvements in operation
fidelity and charge sensitivity are possible by reducing \tmp, using
more complex pulse sequences such as CPMG \cite{Carr1954} and UDD
\cite{Uhrig2007}, and performing operations at larger $J$ and
$dJ/d\epsilon$ to move to a higher frequency portion of the noise
spectrum with potentially lower noise. In particular, because
two-qubit operations in $S$-$T_0$ qubits rely heavily on exchange echo
\cite{Shulman2012}, our data show a path forward for increasing gate
fidelities in these devices and generating higher quality entangled
states. Lastly, the metrological capabilities of the $S$-$T_0$ qubit
may be further improved by harnessing the power of entanglement and
measuring simultaneously with many qubits\cite{Lloyd2011}.

This work is supported through the ARO, \textquotedblleft{}Precision
Quantum Control and Error-Suppressing Quantum Firmware for Robust
Quantum Computing\textquotedblright{} and the IARPA
\textquotedblleft{}Multi-Qubit Coherent Operations (MQCO)
Program\textquotedblright{}.  This work was partially supported by the
US Army Research Office under Contract Number W911NF-11-1-0068. This
work is sponsored by the United States Department of Defence. The
views and conclusions contained in the document are those of the
authors and should not be interpreted as representing the official
policies, either expressly or implied, of the U. S. Government. This
work was performed in part at the Center for Nanoscale Systems (CNS),
a member of the National Nanotechnology Infrastructure Network (NNIN),
which is supported by the National Science Foundation under NSF award
no. ECS-0335765. CNS is part of Harvard University.

\bibliography{bibtex_2012_04_04}{}

\begin{thebibliography}{37}%
\makeatletter
\providecommand \@ifxundefined [1]{%
 \@ifx{#1\undefined}
}%
\providecommand \@ifnum [1]{%
 \ifnum #1\expandafter \@firstoftwo
 \else \expandafter \@secondoftwo
 \fi
}%
\providecommand \@ifx [1]{%
 \ifx #1\expandafter \@firstoftwo
 \else \expandafter \@secondoftwo
 \fi
}%
\providecommand \natexlab [1]{#1}%
\providecommand \enquote  [1]{``#1''}%
\providecommand \bibnamefont  [1]{#1}%
\providecommand \bibfnamefont [1]{#1}%
\providecommand \citenamefont [1]{#1}%
\providecommand \href@noop [0]{\@secondoftwo}%
\providecommand \href [0]{\begingroup \@sanitize@url \@href}%
\providecommand \@href[1]{\@@startlink{#1}\@@href}%
\providecommand \@@href[1]{\endgroup#1\@@endlink}%
\providecommand \@sanitize@url [0]{\catcode `\\12\catcode `\$12\catcode
  `\&12\catcode `\#12\catcode `\^12\catcode `\_12\catcode `\%12\relax}%
\providecommand \@@startlink[1]{}%
\providecommand \@@endlink[0]{}%
\providecommand \url  [0]{\begingroup\@sanitize@url \@url }%
\providecommand \@url [1]{\endgroup\@href {#1}{\urlprefix }}%
\providecommand \urlprefix  [0]{URL }%
\providecommand \Eprint [0]{\href }%
\providecommand \doibase [0]{http://dx.doi.org/}%
\providecommand \selectlanguage [0]{\@gobble}%
\providecommand \bibinfo  [0]{\@secondoftwo}%
\providecommand \bibfield  [0]{\@secondoftwo}%
\providecommand \translation [1]{[#1]}%
\providecommand \BibitemOpen [0]{}%
\providecommand \bibitemStop [0]{}%
\providecommand \bibitemNoStop [0]{.\EOS\space}%
\providecommand \EOS [0]{\spacefactor3000\relax}%
\providecommand \BibitemShut  [1]{\csname bibitem#1\endcsname}%
\let\auto@bib@innerbib\@empty
\bibitem [{\citenamefont {DiVincenzo}(2000)}]{DiVincenzo2000}%
  \BibitemOpen
  \bibfield  {author} {\bibinfo {author} {\bibfnamefont {D.}~\bibnamefont
  {DiVincenzo}},\ }\href@noop {} {\bibfield  {journal} {\bibinfo  {journal}
  {Fortschr. Phys.}\ }\textbf {\bibinfo {volume} {48}},\ \bibinfo {pages} {771}
  (\bibinfo {year} {2000})}\BibitemShut {NoStop}%
\bibitem [{\citenamefont {Chernobrod}\ and\ \citenamefont
  {Brerman}(2005)}]{Chernobrod2005}%
  \BibitemOpen
  \bibfield  {author} {\bibinfo {author} {\bibfnamefont {B.~M.}\ \bibnamefont
  {Chernobrod}}\ and\ \bibinfo {author} {\bibfnamefont {G.~P.}\ \bibnamefont
  {Brerman}},\ }\href {\doibase 10.1063/1.1829373} {\bibfield  {journal}
  {\bibinfo  {journal} {Journal of Applied Physics}\ }\textbf {\bibinfo
  {volume} {97}},\ \bibinfo {pages} {014903} (\bibinfo {year}
  {2005})}\BibitemShut {NoStop}%
\bibitem [{\citenamefont {Maletinsky}\ \emph {et~al.}(2012)\citenamefont
  {Maletinsky}, \citenamefont {Hong}, \citenamefont {Grinolds}, \citenamefont
  {Hausmann}, \citenamefont {Lukin}, \citenamefont {Walsworth}, \citenamefont
  {Loncar},\ and\ \citenamefont {Yacoby}}]{Maletinsky2012}%
  \BibitemOpen
  \bibfield  {author} {\bibinfo {author} {\bibfnamefont {P.}~\bibnamefont
  {Maletinsky}}, \bibinfo {author} {\bibfnamefont {S.}~\bibnamefont {Hong}},
  \bibinfo {author} {\bibfnamefont {M.~S.}\ \bibnamefont {Grinolds}}, \bibinfo
  {author} {\bibfnamefont {B.}~\bibnamefont {Hausmann}}, \bibinfo {author}
  {\bibfnamefont {M.~D.}\ \bibnamefont {Lukin}}, \bibinfo {author}
  {\bibfnamefont {R.~L.}\ \bibnamefont {Walsworth}}, \bibinfo {author}
  {\bibfnamefont {M.}~\bibnamefont {Loncar}}, \ and\ \bibinfo {author}
  {\bibfnamefont {A.}~\bibnamefont {Yacoby}},\ }\href {\doibase
  10.1038/nnano.2012.50} {\bibfield  {journal} {\bibinfo  {journal} {Nature
  Nanotechnology}\ }\textbf {\bibinfo {volume} {7}},\ \bibinfo {pages} {320}
  (\bibinfo {year} {2012})}\BibitemShut {NoStop}%
\bibitem [{\citenamefont {Hanson}\ \emph {et~al.}(2007)\citenamefont {Hanson},
  \citenamefont {Kouwenhoven}, \citenamefont {Petta}, \citenamefont {Tarucha},\
  and\ \citenamefont {Vandersypen}}]{Hanson2007}%
  \BibitemOpen
  \bibfield  {author} {\bibinfo {author} {\bibfnamefont {R.}~\bibnamefont
  {Hanson}}, \bibinfo {author} {\bibfnamefont {L.~P.}\ \bibnamefont
  {Kouwenhoven}}, \bibinfo {author} {\bibfnamefont {J.~R.}\ \bibnamefont
  {Petta}}, \bibinfo {author} {\bibfnamefont {S.}~\bibnamefont {Tarucha}}, \
  and\ \bibinfo {author} {\bibfnamefont {L.~M.~K.}\ \bibnamefont
  {Vandersypen}},\ }\href@noop {} {\bibfield  {journal} {\bibinfo  {journal}
  {Rev. Mod. Phys.}\ }\textbf {\bibinfo {volume} {79}},\ \bibinfo {pages}
  {1217} (\bibinfo {year} {2007})}\BibitemShut {NoStop}%
\bibitem [{\citenamefont {Loss}\ and\ \citenamefont
  {DiVincenzo}(1998)}]{Loss1998}%
  \BibitemOpen
  \bibfield  {author} {\bibinfo {author} {\bibfnamefont {D.}~\bibnamefont
  {Loss}}\ and\ \bibinfo {author} {\bibfnamefont {D.~P.}\ \bibnamefont
  {DiVincenzo}},\ }\href@noop {} {\bibfield  {journal} {\bibinfo  {journal}
  {Phys. Rev. A}\ }\textbf {\bibinfo {volume} {57}},\ \bibinfo {pages} {120}
  (\bibinfo {year} {1998})}\BibitemShut {NoStop}%
\bibitem [{\citenamefont {Koppens}\ \emph {et~al.}(2006)\citenamefont
  {Koppens}, \citenamefont {Buizert}, \citenamefont {Tielrooij}, \citenamefont
  {Vink}, \citenamefont {Nowack}, \citenamefont {Meunier}, \citenamefont
  {Kouwenhoven},\ and\ \citenamefont {Vandersypen}}]{Koppens2006}%
  \BibitemOpen
  \bibfield  {author} {\bibinfo {author} {\bibfnamefont {F.~H.~L.}\
  \bibnamefont {Koppens}}, \bibinfo {author} {\bibfnamefont {C.}~\bibnamefont
  {Buizert}}, \bibinfo {author} {\bibfnamefont {K.~J.}\ \bibnamefont
  {Tielrooij}}, \bibinfo {author} {\bibfnamefont {I.~T.}\ \bibnamefont {Vink}},
  \bibinfo {author} {\bibfnamefont {K.~C.}\ \bibnamefont {Nowack}}, \bibinfo
  {author} {\bibfnamefont {T.}~\bibnamefont {Meunier}}, \bibinfo {author}
  {\bibfnamefont {L.~P.}\ \bibnamefont {Kouwenhoven}}, \ and\ \bibinfo {author}
  {\bibfnamefont {L.~M.~K.}\ \bibnamefont {Vandersypen}},\ }\href@noop {}
  {\bibfield  {journal} {\bibinfo  {journal} {Nature}\ }\textbf {\bibinfo
  {volume} {442}},\ \bibinfo {pages} {766} (\bibinfo {year}
  {2006})}\BibitemShut {NoStop}%
\bibitem [{\citenamefont {Pioro-Ladriere}\ \emph {et~al.}(2008)\citenamefont
  {Pioro-Ladriere}, \citenamefont {Obata}, \citenamefont {Tokura},
  \citenamefont {Shin}, \citenamefont {Kubo}, \citenamefont {Yoshida},
  \citenamefont {Taniyama},\ and\ \citenamefont {Tarucha}}]{Pioro2008}%
  \BibitemOpen
  \bibfield  {author} {\bibinfo {author} {\bibfnamefont {M.}~\bibnamefont
  {Pioro-Ladriere}}, \bibinfo {author} {\bibfnamefont {T.}~\bibnamefont
  {Obata}}, \bibinfo {author} {\bibfnamefont {Y.}~\bibnamefont {Tokura}},
  \bibinfo {author} {\bibfnamefont {Y.~S.}\ \bibnamefont {Shin}}, \bibinfo
  {author} {\bibfnamefont {T.}~\bibnamefont {Kubo}}, \bibinfo {author}
  {\bibfnamefont {K.}~\bibnamefont {Yoshida}}, \bibinfo {author} {\bibfnamefont
  {T.}~\bibnamefont {Taniyama}}, \ and\ \bibinfo {author} {\bibfnamefont
  {S.}~\bibnamefont {Tarucha}},\ }\href@noop {} {\bibfield  {journal} {\bibinfo
   {journal} {Nature Physics}\ }\textbf {\bibinfo {volume} {4}},\ \bibinfo
  {pages} {776} (\bibinfo {year} {2008})}\BibitemShut {NoStop}%
\bibitem [{\citenamefont {Nowack}\ \emph {et~al.}(2007)\citenamefont {Nowack},
  \citenamefont {Koppens}, \citenamefont {Nazarov},\ and\ \citenamefont
  {Vandersypen}}]{Nowack2007}%
  \BibitemOpen
  \bibfield  {author} {\bibinfo {author} {\bibfnamefont {K.~C.}\ \bibnamefont
  {Nowack}}, \bibinfo {author} {\bibfnamefont {F.~H.~L.}\ \bibnamefont
  {Koppens}}, \bibinfo {author} {\bibfnamefont {Y.~V.}\ \bibnamefont
  {Nazarov}}, \ and\ \bibinfo {author} {\bibfnamefont {L.~M.~K.}\ \bibnamefont
  {Vandersypen}},\ }\href@noop {} {\bibfield  {journal} {\bibinfo  {journal}
  {Science}\ }\textbf {\bibinfo {volume} {318}},\ \bibinfo {pages} {1430}
  (\bibinfo {year} {2007})}\BibitemShut {NoStop}%
\bibitem [{\citenamefont {Pioro-Ladriere}\ \emph {et~al.}(2007)\citenamefont
  {Pioro-Ladriere}, \citenamefont {Tokura}, \citenamefont {Kubo},\ and\
  \citenamefont {Tarucha}}]{Pioro2007}%
  \BibitemOpen
  \bibfield  {author} {\bibinfo {author} {\bibfnamefont {M.}~\bibnamefont
  {Pioro-Ladriere}}, \bibinfo {author} {\bibfnamefont {Y.}~\bibnamefont
  {Tokura}}, \bibinfo {author} {\bibfnamefont {T.}~\bibnamefont {Kubo}}, \ and\
  \bibinfo {author} {\bibfnamefont {S.}~\bibnamefont {Tarucha}},\ }\href@noop
  {} {\bibfield  {journal} {\bibinfo  {journal} {Appl. Phys. Lett.}\ }\textbf
  {\bibinfo {volume} {90}},\ \bibinfo {pages} {024105} (\bibinfo {year}
  {2007})}\BibitemShut {NoStop}%
\bibitem [{\citenamefont {Churchill}\ \emph {et~al.}(2009)\citenamefont
  {Churchill}, \citenamefont {Kuemmeth}, \citenamefont {Harlow}, \citenamefont
  {Bestwick}, \citenamefont {Rashba}, \citenamefont {Flensberg}, \citenamefont
  {Stwertka}, \citenamefont {Taychatanapat}, \citenamefont {Watson},\ and\
  \citenamefont {Marcus}}]{Churchill2009}%
  \BibitemOpen
  \bibfield  {author} {\bibinfo {author} {\bibfnamefont {H.~O.~H.}\
  \bibnamefont {Churchill}}, \bibinfo {author} {\bibfnamefont {F.}~\bibnamefont
  {Kuemmeth}}, \bibinfo {author} {\bibfnamefont {J.~W.}\ \bibnamefont
  {Harlow}}, \bibinfo {author} {\bibfnamefont {A.~J.}\ \bibnamefont
  {Bestwick}}, \bibinfo {author} {\bibfnamefont {E.~I.}\ \bibnamefont
  {Rashba}}, \bibinfo {author} {\bibfnamefont {K.}~\bibnamefont {Flensberg}},
  \bibinfo {author} {\bibfnamefont {C.~H.}\ \bibnamefont {Stwertka}}, \bibinfo
  {author} {\bibfnamefont {T.}~\bibnamefont {Taychatanapat}}, \bibinfo {author}
  {\bibfnamefont {S.~K.}\ \bibnamefont {Watson}}, \ and\ \bibinfo {author}
  {\bibfnamefont {C.~M.}\ \bibnamefont {Marcus}},\ }\href {\doibase
  10.1103/PhysRevLett.102.166802} {\bibfield  {journal} {\bibinfo  {journal}
  {Phys. Rev. Lett.}\ }\textbf {\bibinfo {volume} {102}},\ \bibinfo {pages}
  {166802} (\bibinfo {year} {2009})}\BibitemShut {NoStop}%
\bibitem [{\citenamefont {Nadj-Perfe}\ \emph {et~al.}(2010)\citenamefont
  {Nadj-Perfe}, \citenamefont {Frolov}, \citenamefont {Bkkers},\ and\
  \citenamefont {Kouwenhoven}}]{Nadj-Perge2010}%
  \BibitemOpen
  \bibfield  {author} {\bibinfo {author} {\bibfnamefont {S.}~\bibnamefont
  {Nadj-Perfe}}, \bibinfo {author} {\bibfnamefont {S.~M.}\ \bibnamefont
  {Frolov}}, \bibinfo {author} {\bibfnamefont {E.~P. A.~M.}\ \bibnamefont
  {Bkkers}}, \ and\ \bibinfo {author} {\bibfnamefont {L.~P.}\ \bibnamefont
  {Kouwenhoven}},\ }\href@noop {} {\bibfield  {journal} {\bibinfo  {journal}
  {Nature}\ }\textbf {\bibinfo {volume} {468}},\ \bibinfo {pages} {1084}
  (\bibinfo {year} {2010})}\BibitemShut {NoStop}%
\bibitem [{\citenamefont {Levy}(2002)}]{Levy2002}%
  \BibitemOpen
  \bibfield  {author} {\bibinfo {author} {\bibfnamefont {J.}~\bibnamefont
  {Levy}},\ }\href {\doibase 10.1103/PhysRevLett.89.147902} {\bibfield
  {journal} {\bibinfo  {journal} {Phys. Rev. Lett.}\ }\textbf {\bibinfo
  {volume} {89}},\ \bibinfo {pages} {147902} (\bibinfo {year}
  {2002})}\BibitemShut {NoStop}%
\bibitem [{\citenamefont {Petta}\ \emph {et~al.}(2005)\citenamefont {Petta},
  \citenamefont {Johnson}, \citenamefont {Taylor}, \citenamefont {Laird},
  \citenamefont {Yacoby}, \citenamefont {Lukin}, \citenamefont {Marcus},
  \citenamefont {Hanson},\ and\ \citenamefont {Gossard}}]{Petta2005}%
  \BibitemOpen
  \bibfield  {author} {\bibinfo {author} {\bibfnamefont {J.~R.}\ \bibnamefont
  {Petta}}, \bibinfo {author} {\bibfnamefont {A.~C.}\ \bibnamefont {Johnson}},
  \bibinfo {author} {\bibfnamefont {J.~M.}\ \bibnamefont {Taylor}}, \bibinfo
  {author} {\bibfnamefont {E.~A.}\ \bibnamefont {Laird}}, \bibinfo {author}
  {\bibfnamefont {A.}~\bibnamefont {Yacoby}}, \bibinfo {author} {\bibfnamefont
  {M.~D.}\ \bibnamefont {Lukin}}, \bibinfo {author} {\bibfnamefont {C.~M.}\
  \bibnamefont {Marcus}}, \bibinfo {author} {\bibfnamefont {M.~P.}\
  \bibnamefont {Hanson}}, \ and\ \bibinfo {author} {\bibfnamefont {A.~C.}\
  \bibnamefont {Gossard}},\ }\href@noop {} {\bibfield  {journal} {\bibinfo
  {journal} {Science}\ }\textbf {\bibinfo {volume} {309}},\ \bibinfo {pages}
  {2180} (\bibinfo {year} {2005})}\BibitemShut {NoStop}%
\bibitem [{\citenamefont {Taylor}\ \emph {et~al.}(2007)\citenamefont {Taylor},
  \citenamefont {Petta}, \citenamefont {Johnson}, \citenamefont {Yacoby},
  \citenamefont {Marcus},\ and\ \citenamefont {Lukin}}]{Taylor2007}%
  \BibitemOpen
  \bibfield  {author} {\bibinfo {author} {\bibfnamefont {J.~M.}\ \bibnamefont
  {Taylor}}, \bibinfo {author} {\bibfnamefont {J.~R.}\ \bibnamefont {Petta}},
  \bibinfo {author} {\bibfnamefont {A.~C.}\ \bibnamefont {Johnson}}, \bibinfo
  {author} {\bibfnamefont {A.}~\bibnamefont {Yacoby}}, \bibinfo {author}
  {\bibfnamefont {C.~M.}\ \bibnamefont {Marcus}}, \ and\ \bibinfo {author}
  {\bibfnamefont {M.~D.}\ \bibnamefont {Lukin}},\ }\href@noop {} {\bibfield
  {journal} {\bibinfo  {journal} {Phys. Rev. B}\ }\textbf {\bibinfo {volume}
  {76}},\ \bibinfo {pages} {035315} (\bibinfo {year} {2007})}\BibitemShut
  {NoStop}%
\bibitem [{\citenamefont {Taylor}\ \emph {et~al.}(2005)\citenamefont {Taylor},
  \citenamefont {Engel}, \citenamefont {D\"ur}, \citenamefont {Yacoby},
  \citenamefont {Marcus}, \citenamefont {Zoller},\ and\ \citenamefont
  {Lukin}}]{Taylor2005}%
  \BibitemOpen
  \bibfield  {author} {\bibinfo {author} {\bibfnamefont {J.~M.}\ \bibnamefont
  {Taylor}}, \bibinfo {author} {\bibfnamefont {H.-A.}\ \bibnamefont {Engel}},
  \bibinfo {author} {\bibfnamefont {W.}~\bibnamefont {D\"ur}}, \bibinfo
  {author} {\bibfnamefont {A.}~\bibnamefont {Yacoby}}, \bibinfo {author}
  {\bibfnamefont {C.~M.}\ \bibnamefont {Marcus}}, \bibinfo {author}
  {\bibfnamefont {P.}~\bibnamefont {Zoller}}, \ and\ \bibinfo {author}
  {\bibfnamefont {M.~D.}\ \bibnamefont {Lukin}},\ }\href@noop {} {\bibfield
  {journal} {\bibinfo  {journal} {Nature Physics}\ }\textbf {\bibinfo {volume}
  {1}},\ \bibinfo {pages} {177} (\bibinfo {year} {2005})}\BibitemShut {NoStop}%
\bibitem [{\citenamefont {Bluhm}\ \emph {et~al.}(2010)\citenamefont {Bluhm},
  \citenamefont {Foletti}, \citenamefont {Mahalu}, \citenamefont {Umansky},\
  and\ \citenamefont {Yacoby}}]{Bluhm2009}%
  \BibitemOpen
  \bibfield  {author} {\bibinfo {author} {\bibfnamefont {H.}~\bibnamefont
  {Bluhm}}, \bibinfo {author} {\bibfnamefont {S.}~\bibnamefont {Foletti}},
  \bibinfo {author} {\bibfnamefont {D.}~\bibnamefont {Mahalu}}, \bibinfo
  {author} {\bibfnamefont {V.}~\bibnamefont {Umansky}}, \ and\ \bibinfo
  {author} {\bibfnamefont {A.}~\bibnamefont {Yacoby}},\ }\href@noop {}
  {\bibfield  {journal} {\bibinfo  {journal} {Phys. Rev. Lett.}\ }\textbf
  {\bibinfo {volume} {105}},\ \bibinfo {pages} {216803} (\bibinfo {year}
  {2010})}\BibitemShut {NoStop}%
\bibitem [{\citenamefont {Foletti}\ \emph {et~al.}(2009)\citenamefont
  {Foletti}, \citenamefont {Bluhm}, \citenamefont {Mahalu}, \citenamefont
  {Umansky},\ and\ \citenamefont {Yacoby}}]{Foletti2009}%
  \BibitemOpen
  \bibfield  {author} {\bibinfo {author} {\bibfnamefont {S.}~\bibnamefont
  {Foletti}}, \bibinfo {author} {\bibfnamefont {H.}~\bibnamefont {Bluhm}},
  \bibinfo {author} {\bibfnamefont {D.}~\bibnamefont {Mahalu}}, \bibinfo
  {author} {\bibfnamefont {V.}~\bibnamefont {Umansky}}, \ and\ \bibinfo
  {author} {\bibfnamefont {A.}~\bibnamefont {Yacoby}},\ }\href@noop {}
  {\bibfield  {journal} {\bibinfo  {journal} {Nature Physics}\ }\textbf
  {\bibinfo {volume} {5}},\ \bibinfo {pages} {903} (\bibinfo {year}
  {2009})}\BibitemShut {NoStop}%
\bibitem [{\citenamefont {Reilly}\ \emph {et~al.}(2007)\citenamefont {Reilly},
  \citenamefont {Marcus}, \citenamefont {Hanson},\ and\ \citenamefont
  {Gossard}}]{Reilly2007}%
  \BibitemOpen
  \bibfield  {author} {\bibinfo {author} {\bibfnamefont {D.~J.}\ \bibnamefont
  {Reilly}}, \bibinfo {author} {\bibfnamefont {C.~M.}\ \bibnamefont {Marcus}},
  \bibinfo {author} {\bibfnamefont {M.~P.}\ \bibnamefont {Hanson}}, \ and\
  \bibinfo {author} {\bibfnamefont {A.~C.}\ \bibnamefont {Gossard}},\
  }\href@noop {} {\bibfield  {journal} {\bibinfo  {journal} {Appl. Phys.
  Lett.}\ }\textbf {\bibinfo {volume} {91}},\ \bibinfo {pages} {162101}
  (\bibinfo {year} {2007})}\BibitemShut {NoStop}%
\bibitem [{\citenamefont {Barthel}\ \emph {et~al.}(2009)\citenamefont
  {Barthel}, \citenamefont {Reilly}, \citenamefont {Marcus}, \citenamefont
  {Hanson},\ and\ \citenamefont {Gossard}}]{Barthel2009}%
  \BibitemOpen
  \bibfield  {author} {\bibinfo {author} {\bibfnamefont {C.}~\bibnamefont
  {Barthel}}, \bibinfo {author} {\bibfnamefont {D.~J.}\ \bibnamefont {Reilly}},
  \bibinfo {author} {\bibfnamefont {C.~M.}\ \bibnamefont {Marcus}}, \bibinfo
  {author} {\bibfnamefont {M.~P.}\ \bibnamefont {Hanson}}, \ and\ \bibinfo
  {author} {\bibfnamefont {A.~C.}\ \bibnamefont {Gossard}},\ }\href@noop {}
  {\bibfield  {journal} {\bibinfo  {journal} {Phys. Rev. Lett.}\ }\textbf
  {\bibinfo {volume} {103}},\ \bibinfo {pages} {160503} (\bibinfo {year}
  {2009})}\BibitemShut {NoStop}%
\bibitem [{\citenamefont {Bluhm}\ \emph {et~al.}(2011)\citenamefont {Bluhm},
  \citenamefont {Foletti}, \citenamefont {Neder}, \citenamefont {Rudner},
  \citenamefont {Mahalu}, \citenamefont {Umansky},\ and\ \citenamefont
  {Yacoby}}]{Bluhm2011}%
  \BibitemOpen
  \bibfield  {author} {\bibinfo {author} {\bibfnamefont {H.}~\bibnamefont
  {Bluhm}}, \bibinfo {author} {\bibfnamefont {S.}~\bibnamefont {Foletti}},
  \bibinfo {author} {\bibfnamefont {I.}~\bibnamefont {Neder}}, \bibinfo
  {author} {\bibfnamefont {M.~S.}\ \bibnamefont {Rudner}}, \bibinfo {author}
  {\bibfnamefont {D.}~\bibnamefont {Mahalu}}, \bibinfo {author} {\bibfnamefont
  {V.}~\bibnamefont {Umansky}}, \ and\ \bibinfo {author} {\bibfnamefont
  {A.}~\bibnamefont {Yacoby}},\ }\href {\doibase DOI: 10.1038/NPHYS1856}
  {\bibfield  {journal} {\bibinfo  {journal} {Nature Physics}\ }\textbf
  {\bibinfo {volume} {7}},\ \bibinfo {pages} {109} (\bibinfo {year}
  {2011})}\BibitemShut {NoStop}%
\bibitem [{\citenamefont {Barthel}\ \emph {et~al.}(2010)\citenamefont
  {Barthel}, \citenamefont {Medford}, \citenamefont {Marcus}, \citenamefont
  {Hanson},\ and\ \citenamefont {Gossard}}]{Barthel2010b}%
  \BibitemOpen
  \bibfield  {author} {\bibinfo {author} {\bibfnamefont {C.}~\bibnamefont
  {Barthel}}, \bibinfo {author} {\bibfnamefont {J.}~\bibnamefont {Medford}},
  \bibinfo {author} {\bibfnamefont {C.~M.}\ \bibnamefont {Marcus}}, \bibinfo
  {author} {\bibfnamefont {M.~P.}\ \bibnamefont {Hanson}}, \ and\ \bibinfo
  {author} {\bibfnamefont {A.~C.}\ \bibnamefont {Gossard}},\ }\href {\doibase
  10.1103/PhysRevLett.105.266808} {\bibfield  {journal} {\bibinfo  {journal}
  {Phys. Rev. Lett.}\ }\textbf {\bibinfo {volume} {105}},\ \bibinfo {pages}
  {266808} (\bibinfo {year} {2010})}\BibitemShut {NoStop}%
\bibitem [{\citenamefont {Medford}\ \emph {et~al.}(2012)\citenamefont
  {Medford}, \citenamefont {Cywi\ifmmode~\acute{n}\else \'{n}\fi{}ski},
  \citenamefont {Barthel}, \citenamefont {Marcus}, \citenamefont {Hanson},\
  and\ \citenamefont {Gossard}}]{Medford2012}%
  \BibitemOpen
  \bibfield  {author} {\bibinfo {author} {\bibfnamefont {J.}~\bibnamefont
  {Medford}}, \bibinfo {author} {\bibfnamefont {L.}~\bibnamefont
  {Cywi\ifmmode~\acute{n}\else \'{n}\fi{}ski}}, \bibinfo {author}
  {\bibfnamefont {C.}~\bibnamefont {Barthel}}, \bibinfo {author} {\bibfnamefont
  {C.~M.}\ \bibnamefont {Marcus}}, \bibinfo {author} {\bibfnamefont {M.~P.}\
  \bibnamefont {Hanson}}, \ and\ \bibinfo {author} {\bibfnamefont {A.~C.}\
  \bibnamefont {Gossard}},\ }\href {\doibase 10.1103/PhysRevLett.108.086802}
  {\bibfield  {journal} {\bibinfo  {journal} {Phys. Rev. Lett.}\ }\textbf
  {\bibinfo {volume} {108}},\ \bibinfo {pages} {086802} (\bibinfo {year}
  {2012})}\BibitemShut {NoStop}%
\bibitem [{\citenamefont {Laird}\ \emph {et~al.}(2010)\citenamefont {Laird},
  \citenamefont {Taylor}, \citenamefont {DiVincenzo}, \citenamefont {Marcus},
  \citenamefont {Hanson},\ and\ \citenamefont {Gossard}}]{Laird2010}%
  \BibitemOpen
  \bibfield  {author} {\bibinfo {author} {\bibfnamefont {E.~A.}\ \bibnamefont
  {Laird}}, \bibinfo {author} {\bibfnamefont {J.~M.}\ \bibnamefont {Taylor}},
  \bibinfo {author} {\bibfnamefont {D.~P.}\ \bibnamefont {DiVincenzo}},
  \bibinfo {author} {\bibfnamefont {C.~M.}\ \bibnamefont {Marcus}}, \bibinfo
  {author} {\bibfnamefont {M.~P.}\ \bibnamefont {Hanson}}, \ and\ \bibinfo
  {author} {\bibfnamefont {A.~C.}\ \bibnamefont {Gossard}},\ }\href {\doibase
  10.1103/PhysRevB.82.075403} {\bibfield  {journal} {\bibinfo  {journal} {Phys.
  Rev. B}\ }\textbf {\bibinfo {volume} {82}},\ \bibinfo {pages} {075403}
  (\bibinfo {year} {2010})}\BibitemShut {NoStop}%
\bibitem [{\citenamefont {Gaudreau}\ \emph {et~al.}(2011)\citenamefont
  {Gaudreau}, \citenamefont {Granger}, \citenamefont {Kam}, \citenamefont
  {Aers}, \citenamefont {andP. Zawadzki}, \citenamefont {Pioro-Ladrière},
  \citenamefont {Wasilewski},\ and\ \citenamefont {Sachrajda}}]{Gaudreau2011}%
  \BibitemOpen
  \bibfield  {author} {\bibinfo {author} {\bibfnamefont {L.}~\bibnamefont
  {Gaudreau}}, \bibinfo {author} {\bibfnamefont {G.}~\bibnamefont {Granger}},
  \bibinfo {author} {\bibfnamefont {A.}~\bibnamefont {Kam}}, \bibinfo {author}
  {\bibfnamefont {G.~C.}\ \bibnamefont {Aers}}, \bibinfo {author}
  {\bibfnamefont {S.~A.~S.}\ \bibnamefont {andP. Zawadzki}}, \bibinfo {author}
  {\bibfnamefont {M.}~\bibnamefont {Pioro-Ladrière}}, \bibinfo {author}
  {\bibfnamefont {Z.~R.}\ \bibnamefont {Wasilewski}}, \ and\ \bibinfo {author}
  {\bibfnamefont {A.~S.}\ \bibnamefont {Sachrajda}},\ }\href {\doibase
  doi:10.1038/nphys2149} {\bibfield  {journal} {\bibinfo  {journal} {Nature
  Physics}\ }\textbf {\bibinfo {volume} {8}},\ \bibinfo {pages} {54} (\bibinfo
  {year} {2011})}\BibitemShut {NoStop}%
\bibitem [{\citenamefont {Shulman}\ \emph {et~al.}(2012)\citenamefont
  {Shulman}, \citenamefont {Dial}, \citenamefont {Harvey}, \citenamefont
  {Bluhm}, \citenamefont {Umansky},\ and\ \citenamefont
  {Yacoby}}]{Shulman2012}%
  \BibitemOpen
  \bibfield  {author} {\bibinfo {author} {\bibfnamefont {M.~D.}\ \bibnamefont
  {Shulman}}, \bibinfo {author} {\bibfnamefont {O.~E.}\ \bibnamefont {Dial}},
  \bibinfo {author} {\bibfnamefont {S.~P.}\ \bibnamefont {Harvey}}, \bibinfo
  {author} {\bibfnamefont {H.}~\bibnamefont {Bluhm}}, \bibinfo {author}
  {\bibfnamefont {V.}~\bibnamefont {Umansky}}, \ and\ \bibinfo {author}
  {\bibfnamefont {A.}~\bibnamefont {Yacoby}},\ }\href {\doibase
  10.1126/science.1217692} {\bibfield  {journal} {\bibinfo  {journal}
  {Science}\ }\textbf {\bibinfo {volume} {336}},\ \bibinfo {pages} {202}
  (\bibinfo {year} {2012})}\BibitemShut {NoStop}%
\bibitem [{\citenamefont {van Weperen}\ \emph {et~al.}(2011)\citenamefont {van
  Weperen}, \citenamefont {Armstrong}, \citenamefont {Laird}, \citenamefont
  {Medford}, \citenamefont {Marcus}, \citenamefont {Hanson},\ and\
  \citenamefont {Gossard}}]{VanWeperen2011}%
  \BibitemOpen
  \bibfield  {author} {\bibinfo {author} {\bibfnamefont {I.}~\bibnamefont {van
  Weperen}}, \bibinfo {author} {\bibfnamefont {B.~D.}\ \bibnamefont
  {Armstrong}}, \bibinfo {author} {\bibfnamefont {E.~A.}\ \bibnamefont
  {Laird}}, \bibinfo {author} {\bibfnamefont {J.}~\bibnamefont {Medford}},
  \bibinfo {author} {\bibfnamefont {C.~M.}\ \bibnamefont {Marcus}}, \bibinfo
  {author} {\bibfnamefont {M.~P.}\ \bibnamefont {Hanson}}, \ and\ \bibinfo
  {author} {\bibfnamefont {A.~C.}\ \bibnamefont {Gossard}},\ }\href {\doibase
  10.1103/PhysRevLett.107.030506} {\bibfield  {journal} {\bibinfo  {journal}
  {Phys. Rev. Lett.}\ }\textbf {\bibinfo {volume} {107}},\ \bibinfo {pages}
  {030506} (\bibinfo {year} {2011})}\BibitemShut {NoStop}%
\bibitem [{\citenamefont {Brunner}\ \emph {et~al.}(2011)\citenamefont
  {Brunner}, \citenamefont {Shin}, \citenamefont {Obata}, \citenamefont
  {Pioro-Ladri\`ere}, \citenamefont {Kubo}, \citenamefont {Yoshida},
  \citenamefont {Taniyama}, \citenamefont {Tokura},\ and\ \citenamefont
  {Tarucha}}]{Brunner2011}%
  \BibitemOpen
  \bibfield  {author} {\bibinfo {author} {\bibfnamefont {R.}~\bibnamefont
  {Brunner}}, \bibinfo {author} {\bibfnamefont {Y.-S.}\ \bibnamefont {Shin}},
  \bibinfo {author} {\bibfnamefont {T.}~\bibnamefont {Obata}}, \bibinfo
  {author} {\bibfnamefont {M.}~\bibnamefont {Pioro-Ladri\`ere}}, \bibinfo
  {author} {\bibfnamefont {T.}~\bibnamefont {Kubo}}, \bibinfo {author}
  {\bibfnamefont {K.}~\bibnamefont {Yoshida}}, \bibinfo {author} {\bibfnamefont
  {T.}~\bibnamefont {Taniyama}}, \bibinfo {author} {\bibfnamefont
  {Y.}~\bibnamefont {Tokura}}, \ and\ \bibinfo {author} {\bibfnamefont
  {S.}~\bibnamefont {Tarucha}},\ }\href {\doibase
  10.1103/PhysRevLett.107.146801} {\bibfield  {journal} {\bibinfo  {journal}
  {Phys. Rev. Lett.}\ }\textbf {\bibinfo {volume} {107}},\ \bibinfo {pages}
  {146801} (\bibinfo {year} {2011})}\BibitemShut {NoStop}%
\bibitem [{\citenamefont {Nowack}\ \emph {et~al.}(2011)\citenamefont {Nowack},
  \citenamefont {Shafiei}, \citenamefont {Laforest}, \citenamefont
  {Prawiroatmodjo}, \citenamefont {Schreiber}, \citenamefont {Reichl},
  \citenamefont {Wegscheider},\ and\ \citenamefont {Vandersypen}}]{Nowack2011}%
  \BibitemOpen
  \bibfield  {author} {\bibinfo {author} {\bibfnamefont {K.~C.}\ \bibnamefont
  {Nowack}}, \bibinfo {author} {\bibfnamefont {M.}~\bibnamefont {Shafiei}},
  \bibinfo {author} {\bibfnamefont {M.}~\bibnamefont {Laforest}}, \bibinfo
  {author} {\bibfnamefont {G.~E. D.~K.}\ \bibnamefont {Prawiroatmodjo}},
  \bibinfo {author} {\bibfnamefont {L.~R.}\ \bibnamefont {Schreiber}}, \bibinfo
  {author} {\bibfnamefont {C.}~\bibnamefont {Reichl}}, \bibinfo {author}
  {\bibfnamefont {W.}~\bibnamefont {Wegscheider}}, \ and\ \bibinfo {author}
  {\bibfnamefont {L.~M.~K.}\ \bibnamefont {Vandersypen}},\ }\href {\doibase
  10.1126/science.1209524} {\bibfield  {journal} {\bibinfo  {journal}
  {Science}\ }\textbf {\bibinfo {volume} {333}},\ \bibinfo {pages} {1269}
  (\bibinfo {year} {2011})}\BibitemShut {NoStop}%
\bibitem [{\citenamefont {Culcer}\ \emph {et~al.}(2009)\citenamefont {Culcer},
  \citenamefont {Hu},\ and\ \citenamefont {Sarma}}]{Culcer2009}%
  \BibitemOpen
  \bibfield  {author} {\bibinfo {author} {\bibfnamefont {D.}~\bibnamefont
  {Culcer}}, \bibinfo {author} {\bibfnamefont {X.}~\bibnamefont {Hu}}, \ and\
  \bibinfo {author} {\bibfnamefont {S.~D.}\ \bibnamefont {Sarma}},\ }\href@noop
  {} {\bibfield  {journal} {\bibinfo  {journal} {Appl. Phys. Lett}\ }\textbf
  {\bibinfo {volume} {95}},\ \bibinfo {pages} {073102} (\bibinfo {year}
  {2009})}\BibitemShut {NoStop}%
\bibitem [{\citenamefont {Maune}\ \emph {et~al.}(2011)\citenamefont {Maune},
  \citenamefont {Borselli}, \citenamefont {Huang}, \citenamefont {Ladd},
  \citenamefont {Deelman}, \citenamefont {Holabird}, \citenamefont {Kiselev},
  \citenamefont {Alvarado-Rodriguez}, \citenamefont {Ross}, \citenamefont
  {Schmitz}, \citenamefont {Sokolich}, \citenamefont {Watson}, \citenamefont
  {Gyure},\ and\ \citenamefont {Hunter}}]{Maune2011}%
  \BibitemOpen
  \bibfield  {author} {\bibinfo {author} {\bibfnamefont {B.~M.}\ \bibnamefont
  {Maune}}, \bibinfo {author} {\bibfnamefont {M.~G.}\ \bibnamefont {Borselli}},
  \bibinfo {author} {\bibfnamefont {B.}~\bibnamefont {Huang}}, \bibinfo
  {author} {\bibfnamefont {T.~D.}\ \bibnamefont {Ladd}}, \bibinfo {author}
  {\bibfnamefont {P.~W.}\ \bibnamefont {Deelman}}, \bibinfo {author}
  {\bibfnamefont {K.~S.}\ \bibnamefont {Holabird}}, \bibinfo {author}
  {\bibfnamefont {A.~A.}\ \bibnamefont {Kiselev}}, \bibinfo {author}
  {\bibfnamefont {I.}~\bibnamefont {Alvarado-Rodriguez}}, \bibinfo {author}
  {\bibfnamefont {R.~S.}\ \bibnamefont {Ross}}, \bibinfo {author}
  {\bibfnamefont {A.~E.}\ \bibnamefont {Schmitz}}, \bibinfo {author}
  {\bibfnamefont {M.}~\bibnamefont {Sokolich}}, \bibinfo {author}
  {\bibfnamefont {C.~A.}\ \bibnamefont {Watson}}, \bibinfo {author}
  {\bibfnamefont {M.~F.}\ \bibnamefont {Gyure}}, \ and\ \bibinfo {author}
  {\bibfnamefont {A.~T.}\ \bibnamefont {Hunter}},\ }\href {\doibase
  10.1038/nature10707} {\bibfield  {journal} {\bibinfo  {journal} {Nature}\
  }\textbf {\bibinfo {volume} {481}},\ \bibinfo {pages} {344} (\bibinfo {year}
  {2011})}\BibitemShut {NoStop}%
\bibitem [{\citenamefont {Cywinski}\ \emph {et~al.}(2008)\citenamefont
  {Cywinski}, \citenamefont {Lutchyn}, \citenamefont {Nave},\ and\
  \citenamefont {Sarma}}]{Cywinski2008B}%
  \BibitemOpen
  \bibfield  {author} {\bibinfo {author} {\bibfnamefont {L.}~\bibnamefont
  {Cywinski}}, \bibinfo {author} {\bibfnamefont {R.~M.}\ \bibnamefont
  {Lutchyn}}, \bibinfo {author} {\bibfnamefont {C.~P.}\ \bibnamefont {Nave}}, \
  and\ \bibinfo {author} {\bibfnamefont {S.~D.}\ \bibnamefont {Sarma}},\
  }\href@noop {} {\bibfield  {journal} {\bibinfo  {journal} {Phys. Rev. B}\
  }\textbf {\bibinfo {volume} {77}},\ \bibinfo {pages} {174509} (\bibinfo
  {year} {2008})}\BibitemShut {NoStop}%
\bibitem [{\citenamefont {Hahn}(1950)}]{Hahn1950}%
  \BibitemOpen
  \bibfield  {author} {\bibinfo {author} {\bibfnamefont {E.~L.}\ \bibnamefont
  {Hahn}},\ }\href@noop {} {\bibfield  {journal} {\bibinfo  {journal} {Phys.
  Rev.}\ }\textbf {\bibinfo {volume} {80}},\ \bibinfo {pages} {580} (\bibinfo
  {year} {1950})}\BibitemShut {NoStop}%
\bibitem [{\citenamefont {Maze}\ \emph {et~al.}(2008)\citenamefont {Maze},
  \citenamefont {Stanwix}, \citenamefont {Hodges}, \citenamefont {Hong},
  \citenamefont {Taylor}, \citenamefont {Cappellaro}, \citenamefont {Jian},
  \citenamefont {Dutt}, \citenamefont {Togan}, \citenamefont {Zibrov},
  \citenamefont {Yacoby}, \citenamefont {Walsworth},\ and\ \citenamefont
  {Lukin}}]{Maze2008}%
  \BibitemOpen
  \bibfield  {author} {\bibinfo {author} {\bibfnamefont {J.~R.}\ \bibnamefont
  {Maze}}, \bibinfo {author} {\bibfnamefont {P.~L.}\ \bibnamefont {Stanwix}},
  \bibinfo {author} {\bibfnamefont {J.~S.}\ \bibnamefont {Hodges}}, \bibinfo
  {author} {\bibfnamefont {S.}~\bibnamefont {Hong}}, \bibinfo {author}
  {\bibfnamefont {J.~M.}\ \bibnamefont {Taylor}}, \bibinfo {author}
  {\bibfnamefont {P.}~\bibnamefont {Cappellaro}}, \bibinfo {author}
  {\bibfnamefont {L.}~\bibnamefont {Jian}}, \bibinfo {author} {\bibfnamefont
  {M.~V.~G.}\ \bibnamefont {Dutt}}, \bibinfo {author} {\bibfnamefont
  {E.}~\bibnamefont {Togan}}, \bibinfo {author} {\bibfnamefont {A.~S.}\
  \bibnamefont {Zibrov}}, \bibinfo {author} {\bibfnamefont {A.}~\bibnamefont
  {Yacoby}}, \bibinfo {author} {\bibfnamefont {R.~L.}\ \bibnamefont
  {Walsworth}}, \ and\ \bibinfo {author} {\bibfnamefont {M.~D.}\ \bibnamefont
  {Lukin}},\ }\href {\doibase 10.1038/nature07279} {\bibfield  {journal}
  {\bibinfo  {journal} {Nature}\ }\textbf {\bibinfo {volume} {455}},\ \bibinfo
  {pages} {644} (\bibinfo {year} {2008})}\BibitemShut {NoStop}%
\bibitem [{\citenamefont {Carr}\ and\ \citenamefont
  {Purcell}(1954)}]{Carr1954}%
  \BibitemOpen
  \bibfield  {author} {\bibinfo {author} {\bibfnamefont {H.~Y.}\ \bibnamefont
  {Carr}}\ and\ \bibinfo {author} {\bibfnamefont {E.~M.}\ \bibnamefont
  {Purcell}},\ }\href@noop {} {\bibfield  {journal} {\bibinfo  {journal} {Phys.
  Rev.}\ }\textbf {\bibinfo {volume} {94}},\ \bibinfo {pages} {630} (\bibinfo
  {year} {1954})}\BibitemShut {NoStop}%
\bibitem [{\citenamefont {Uhrig}(2007)}]{Uhrig2007}%
  \BibitemOpen
  \bibfield  {author} {\bibinfo {author} {\bibfnamefont {G.~S.}\ \bibnamefont
  {Uhrig}},\ }\href@noop {} {\bibfield  {journal} {\bibinfo  {journal} {Phys.
  Rev. Lett.}\ }\textbf {\bibinfo {volume} {98}},\ \bibinfo {pages} {100504}
  (\bibinfo {year} {2007})}\BibitemShut {NoStop}%
\bibitem [{\citenamefont {Devoret}\ and\ \citenamefont
  {Schoelkopf}(2000)}]{Devoret2000}%
  \BibitemOpen
  \bibfield  {author} {\bibinfo {author} {\bibfnamefont {M.~H.}\ \bibnamefont
  {Devoret}}\ and\ \bibinfo {author} {\bibfnamefont {R.~J.}\ \bibnamefont
  {Schoelkopf}},\ }\href@noop {} {\bibfield  {journal} {\bibinfo  {journal}
  {Nature}\ }\textbf {\bibinfo {volume} {406}},\ \bibinfo {pages} {1039}
  (\bibinfo {year} {2000})}\BibitemShut {NoStop}%
\bibitem [{\citenamefont {Giovannetti}\ \emph {et~al.}(2011)\citenamefont
  {Giovannetti}, \citenamefont {Lloyd},\ and\ \citenamefont
  {Maccone}}]{Lloyd2011}%
  \BibitemOpen
  \bibfield  {author} {\bibinfo {author} {\bibfnamefont {V.}~\bibnamefont
  {Giovannetti}}, \bibinfo {author} {\bibfnamefont {S.}~\bibnamefont {Lloyd}},
  \ and\ \bibinfo {author} {\bibfnamefont {L.}~\bibnamefont {Maccone}},\ }\href
  {\doibase 10.1038/nphoton.2011.35} {\bibfield  {journal} {\bibinfo  {journal}
  {Nature Photonics}\ }\textbf {\bibinfo {volume} {5}},\ \bibinfo {pages} {222}
  (\bibinfo {year} {2011})}\BibitemShut {NoStop}%
\end{thebibliography}%


\end{document}


\title{Supplemental Material for {} ``Electrometry Using Coherent
Exchange Oscillations in a Singlet-Triplet Qubit''}
\date{\today}
\author{O. E. Dial}
\author{M. D. Shulman}
\author{S. P. Harvey}
\affiliation{Department of Physics, Harvard University, Cambridge, MA, 02138, USA}
\author{H. Bluhm}
\affiliation{Department of Physics, Harvard University, Cambridge, MA, 02138, USA}
\affiliation{Current Address: 2nd Institute of Physics C, RWTH Aachen University, 52074 Aachen, Germany}
\author{V. Umansky}
\affiliation{Braun Center for Submicron Research, Department
of Condensed Matter Physics, Weizmann Institude of Science, Rehovot
76100 Israel}
\author{A. Yacoby}
\affiliation{Department of Physics, Harvard University, Cambridge, MA, 02138, USA}
\pacs{85.35.Be 03.67.-a 07.50.Ls 76.60.Lz 07.50.Hp}
\maketitle

\section{Measuring $J(\epsilon)$}

\begin{figure}
\resizebox{\columnwidth}{!}{\includegraphics{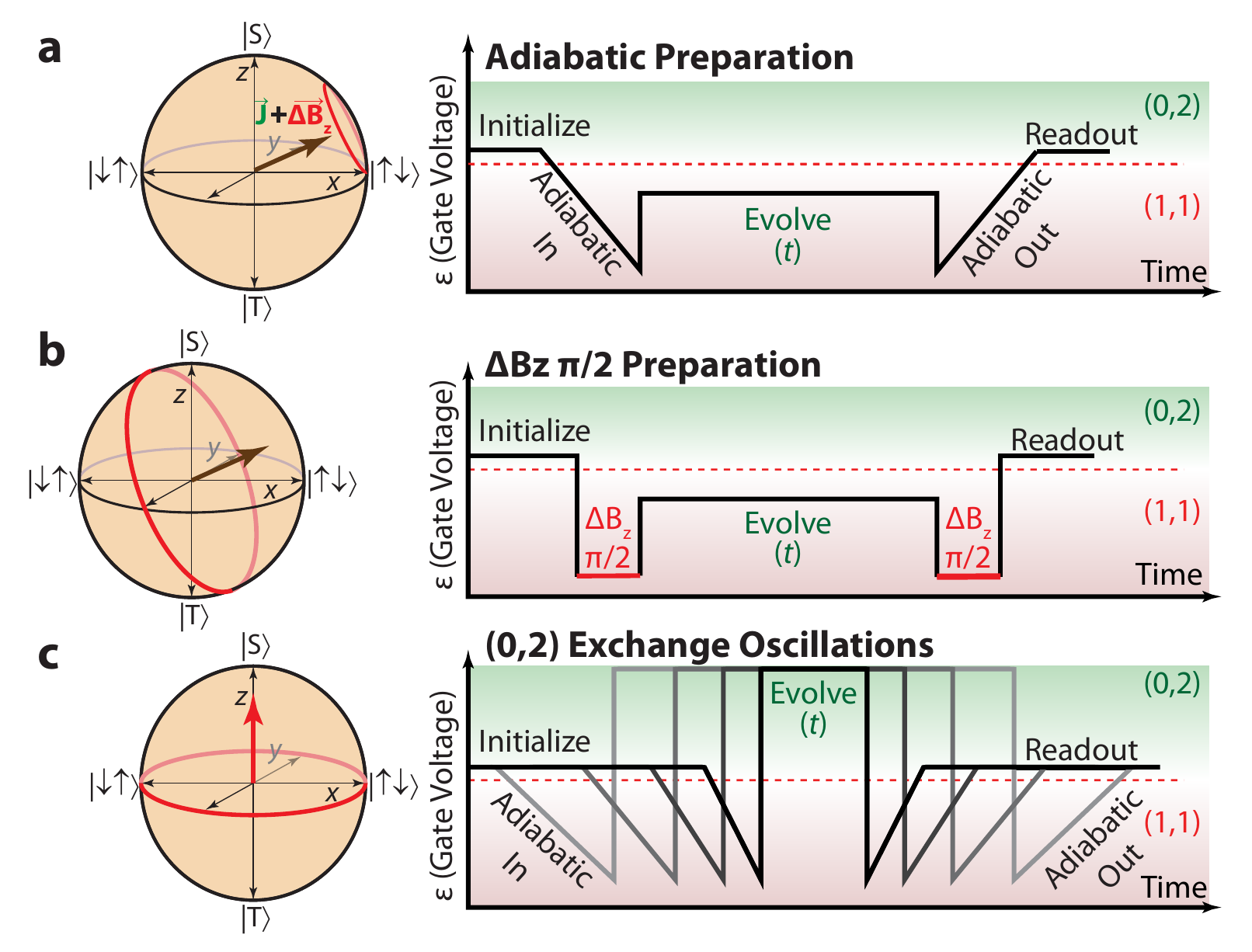}}
\caption{\textbf{a}, The pulse sequence traditionally used to measure
  exchange oscillations in $(1,1)$ employs adiabatic preparation and
  readout but suffers reduced visibility when $J$ becomes comparable
  to $\Delta{}B_{Z}$. Red circles show the path of the state vector on
  the Bloch sphere during the exchange rotation for each sequence.
  \textbf{b}, We use a feedback-controlled nuclear gradient to prepare
  and readout on the $y$-axis, which provides good visibility for all
  values of $J$. \textbf{c}, The pulse sequence used to measure
  exchange oscillations in $(0,2)$ uses adiabatic preparation and
  readout and ``stretches'' the pulse by changing the waveform
  generator clock rate.
\label{pulseSequences}
}
\end{figure}

Previously$^{13,30}$, exchange oscillations were measured inside of
(1,1) by preparing a $\ket{S}$, adiabatically turning $J$ off to
prepare a state on the $x$-axis of the Bloch sphere and then turning
$J$ on again to allow the qubit state to rotate around the $z$-axis.
Finally, to map the $x$-axis of the Bloch sphere back to the $z$-axis
for readout $J$ is turned off suddenly and then adiabatically turned
back on again (\figref{pulseSequences}{a}).  This allows measurements
without a stabilized nuclear gradient. For $J \gg \Delta B_Z$, $J$
drives rotations around the $z$-axis leading to oscillations with
large visibility, but as $J$ becomes comparable to $\Delta B_Z/2\pi$,
$\sim$30 MHz in this work, the axis of rotation tilts towards the
$x$-axis and the visibility of exchange oscillations becomes small
(Bloch sphere in \figref{pulseSequences}{a}).  In order to measure the
value of $J$ when it is small, we use a modified pulse sequence where
we load the qubit in $\ket{S}$, perform a $\pi/2-$pulse around the
$x$-axis to prepare a state on the $y$-axis, then turn on $J$ for a
time $t$, and finally perform a $\pi/2-$pulse to project onto
$\ket{S}$ (\figref{pulseSequences}{b}). We use $\pi/2-$pulses around
the $x$-axis because it guarantees good contrast by preparing the
qubit in a state that is perpendicular to the axis of rotation for all
values of $J$. We note that this sequence requires a stabilized
hyperfine gradient.  The observed oscillation is driven by a
combination of $J$ and $\Delta B_Z$ In order to extract only the
component of the splitting due to the exchange interaction, we measure
the magnetic field gradient $\Delta{}B_{Z}$ with each measurement of
the splitting and calculate
$J(\epsilon)=\sqrt{J_{tot}^{2}-(\Delta{}B_{Z})^{2}}$ where $J_{tot}$
is the observed total frequency.

With an AWG clock frequency of 12 GHz, we can directly measure $J_{tot}$ up
to a few GHz.  To measure faster oscillations
we change the clock rate of the waveform generator,
which stretches the pulse, changing the evolution time in small
increments.  We use adiabatic preparation and readout instead of
$\pi/2-$pulses around the $x$-axis because they do not change when the
pulse is stretched by changing the waveform generator clock and do not
require a stabilized nuclear gradient
(\figref{pulseSequences}{c}). Using this method, we are able to
resolve exchange oscillations exceeding 30GHz.  The axis error (see
below) associated with using an adiabatic preparation is negligible
when measuring such large values of exchange since the angle between
the $z$-axis and axis of revolution $\theta \sim .01-.001$ radians. We
note that for these measurements we use Minicircuits SBLP-467+ low pass
filters on the fast detuning gates to prevent Zener tunneling when crossing
the $(0,2)$-$(1,1)$ charge transition.  

For an exchange echo experiment, there are two ways in
which axis errors lead to a reduction in visibility.
  As in FID, if an adiabatic preparation is used, the
axis of rotation is not perpendicular to the prepared state, and we
avoid this problem by using $\pi/2$-pulses
around the $x$-axis to prepare and read out the qubit.
The second axis error arises from the fact that the axis around which
we apply the $\pi$-pulse for decoupling is not perpendicular to the
axis of rotation. It is possible to mitigate this adverse effect by
using a $\pi$ rotation around the $y$-axis, although for simplicity in
this work we do not.  The
expected visibility as a function of $\theta$ for different sequences
is summarized in Table S1.

\begin{table}
\center
\begin{tabular}{|l | r|}
\hline
Rotation Sequence & Ideal Visibility\\
\hline
FID, adiabatic prep & $\cos^{2}(\theta)$\\
FID, $x$-$\pi/2$-prep & $1$ \\
Echo, adiabatic prep, $x$-$\pi$ & $\cos^{4}(\theta)$\\
Echo, adiabatic prep, $y$-$\pi$ & $\cos^{2}(\theta)$\\
Echo, $x$-$\pi/2$ prep, $x$-$\pi$ & $\cos^{2}(\theta)$\\
Echo, $x$-$\pi/2$ prep, $y$-$\pi$ & 1 \\
\hline
\end{tabular}
\caption{The visibility loss associated with the
  different preparations and $\pi$-pulses.
\label{axiserror}
}
\end{table}

Pulse risetimes in the experimental setup cause the frequency of
exchange oscillations to slowly chirp up to a steady value over
several nanoseconds. This chirp causes us to underestimate
$J(\epsilon)$ in the region where $T_{2}^{*}$ is comparable to the
time required for $J$ to saturate. Consequently, we are unable to
assign unbiased values for $J(\epsilon)$ or $T_{2}^{*}$ in the regions
of the largest $\frac{dJ}{d\epsilon}$, in this work roughly $-1
\mathrm{mV} < \epsilon < 0 mV$.  Unlike some previous works,
$\epsilon$=0 is the detuning at which we measure the state of the
qubit and not the degeneracy point between $(0,2)$ and $(1,1)$.

\section{Saturation of $T_{2}^{*}$}
When $J(\epsilon)$ is small compared to $\Delta{}B_{Z}$,
$T_{2}^{*}$ no longer scales with $dJ/d\epsilon$, but rather saturates
at $T_{2,nuclear}^{*}$ which is limited by the quality of the nuclear
feedback.  Geometric considerations give
\begin{equation}
\left(T_{2}^{*}\right)^{-2} = \left(\frac{J}{J_{tot}}\right)^2
\left(T_{2,voltage}^{*}\right)^{-2}+ \left(\frac{\Delta
  B_{Z}}{J_{tot}}\right)^2 \left(T_{2,nuclear}^{*}\right)^{-2}
\end{equation}
where $J$ is the exchange splitting,
$J_{tot}=\sqrt{J^2+\Delta{}B_{Z}^2}$ is the observed frequency, and
$T_{2,voltage}^{*}$ and $T_{2,nuclear}^{*}$ are the independent
coherence times for $J$ and $\Delta{}B_{Z}$ noise, respectively. We
use this form to plot the expected $T_{2}^{*}$ and
note that this fit has no free parameters since we measure
$J$, $\Delta{}B_{Z}$, and $T_{2,nuclear}^{*}$.

\section{$T_{2}^{*}$   and $ T_{meas}$}
\begin{figure}
\resizebox{\columnwidth}{!}{\includegraphics{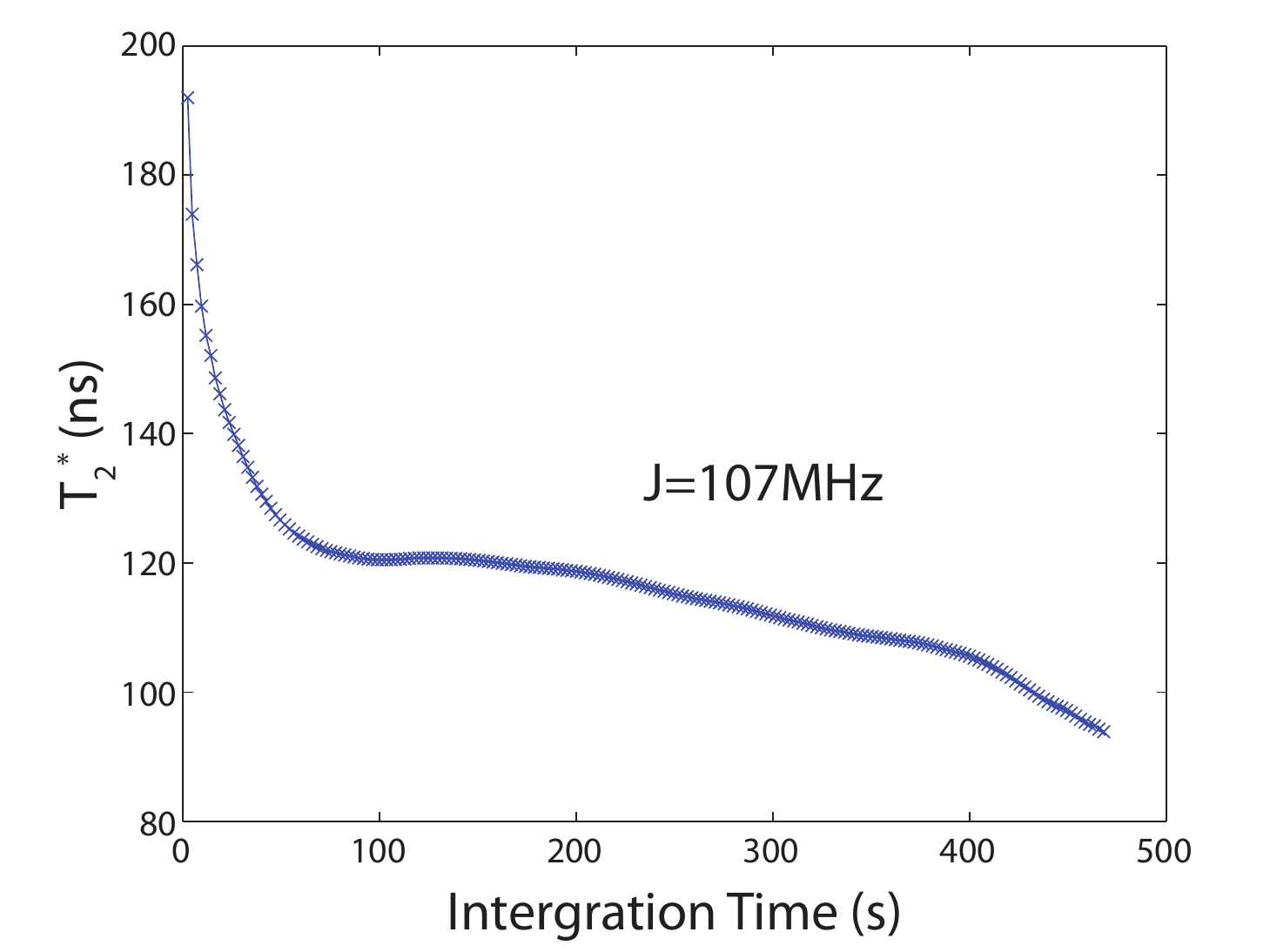}}
\caption{As we measure exchange oscillations for longer,
  increasingly low frequency noise contributes to dephasing and
   $T_{2}^{*}$ decreases.
\label{omegaIR}
}
\end{figure}

Nominally, an FID measurement is sensitive to charge noise from DC to
$J$.  However, any given experiment only runs over a period of time
$T_{meas}$, and this introduces and additional low frequency cutoff;
noise with a frequency below $1/T_{meas}$ will appear as an offset to
$J$ and will not contribute to dephasing. This is particularly
relevant for $1/f$ noise, which diverges at DC.  In general, as the
measurement time increases, the low-frequency cutoff decreases and
the observed $T_2^{*}$ becomes smaller (\figref{omegaIR}).  This makes
it essential to normalize for total measurement time in comparing
results between different groups and experiments.

\section{Fitting Echo Data}
In a typical Hahn echo experiment, we measure the probability of finding a
singlet as a function of total evolution time $\tau$ and change in length of
the second exchange interval $\delta t$.  In order to
self-consistently and reliably fit the data, we simultaneously
fit all the echo data for a given value of $\epsilon$.  For the power-law
noise model, we use the form
\begin{align*}
P_S(\tau,\delta{}t) &= A_0 \cos\left[\omega(\tau) \delta{}t +
  \phi(\tau)\right] \times \\ & \exp\left[-(\tau/T_2)^{\beta+1}
  -\left((\delta{}t-t_0(\tau))/T_2^*\right)^2\right]
\end{align*}
Allowing $\omega$, $\phi$, and the echo center, $t_0$, to vary with
$\tau$ allows us to fit in the presence of finite pulse rise times
which gives small phase and frequency shifts as a function of $\tau$.
Fitting the overall envelope globally through $A_0$, $\beta$,
$T_2$, and $T_2^*$ across many values of $\tau$ greatly reduces the
probability of encountering local minima during fitting.

 In some instances if the $\pi/2$ and $\pi$ pulses are not tuned
 perfectly, there may be some ``un-echoed'' signal at short total
 evolution times such that $T_2^* \gtrsim \tau$.  In this case, we add
 an additional term to $P_S(\tau,t)$
\begin{equation}
P_S'(\tau,\delta{}t) = P_S(\tau,\delta{}t) + A_1 \exp\left[-(\tau/T_2^*)^2\right]
\cos\left[\omega(\tau)\delta{}t + \theta(\tau)\right]
\end{equation}
Here, the new $\tau$-dependent fit paraemter $\theta(\tau)$ accounts
for the phase shift of the unechoed signal due to pulse rise
times$^{25}$, while $A_1$ accounts for the magnitude of the un-echoed
signal.

\section{Dephasing from Other Fluctuating Parameters}
There is no reason, a priori, to assume that the dephasing present in
FID oscillations is due only to fluctuations in $\epsilon$. For example,
in a simple avoided crossing model we expect $J$ to depend on
the tunnel coupling, $T_C$, between the QDs. However, the observed
linear dependence of $T_{2}^{*}$ on $(dJ/d\epsilon)^{-1}$ strongly suggests
that fluctuations in $\epsilon$ are responsible for the observed
dephasing.  For $T_C$ to dominate dephasing, we would need $dJ/dT_C \propto dJ/d\epsilon$ for
all $\epsilon$, a condition which is not true for the simple avoided crossing model
and would generally seem unlikely.  The same argument can be repeated for any other parameter
that couples to $J$.

\section{Extracting Noise Power}
As per Cywinski \emph{et al.}$^{31}$, we calculate the amplitude of
qubit oscillations by integrating the noise spectrum multiplied by an
experiment-specific filter function over all frequencies. For example,
for a Hahn echo experiment in the presence of power-law noise
$S_{\phi}(\omega)=\frac{S}{\omega^{\beta}}$ we find
\begin{equation}
\left(T_{2}^{echo}\right)^{1+\beta}=\frac{\pi}{2^{-\beta}(-2+2^{\beta})S\Gamma{}(-1-\beta)\sin(\pi \beta/2)}
\end{equation}
We note that
$S_{\phi}=\frac{1}{2}S_{\epsilon}(\frac{dJ}{d\epsilon})^{2}$, where
$S_{\epsilon}$ is the $\epsilon$ noise spectrum for positive
frequencies only.  From this we find that
$(T_{2}^{echo})^{-(\beta+1)/2} \propto \frac{dJ}{d\epsilon}$.

For FID experiments, which are dephased by low frequency noise, we
assume the noise is quasistatic (much slower than $1/T_2^*$) and
normally distributed with variance $\sigma^2$. For a given experiment
we find the probability of measuring a singlet
$P(S)=\int_{-\infty}^{\infty}d\omega e^{-(\omega -
  \omega_0)^2/2\sigma^2} e^{i\omega t}$, which can be integrated to
obtain $P(S)= e^{i\omega_0 t}e^{-t^2 \sigma^{2}/2}$. From here we find
$T_2^{*2}=\frac{2}{\sigma^2}=\frac{2}{\epsilon_{RMS}^{2}(2 \pi dJ/d\epsilon)^2}$ where $J$ is in Hertz.

Alternatively, we can consider dephasing from $1/f$ noise during FID
experiments, which leads to nearly Gaussian decay. More precisely, for
noise described by a noise power $S_\epsilon(\omega) = S_1/\omega$,
the decay envelope of FID oscillations is given by $\exp
[-\frac{S_1}{4\pi} \left(\frac{dJ}{d\epsilon}\right)^{2}
t^{2}ln\frac{T_{meas}}{2\pi{}t}]$ where the total measurement time
$T_{meas}\approx{}5 $ minutes provides a low frequeny cutoff to the
noise sensitivity$^{26}$ (see above).  For these measurements,
relevant values of $t$ for decay are of order 100 nanoseconds, so we
neglect the weak $t$ dependence of the logarithmic term and
approximate it by 20.  We fit the decay envelope to the form
$\exp{\left[-(t/T_2^*)^2\right]}$, where
$T_{2}^{*2}=\frac{4\pi}{S_{1}(dJ/d\epsilon)^2ln\frac{T_{meas}}{2\pi{}t}}$
and extract $\sqrt{S_\epsilon} = 2 \mu\mathrm{V}/
\sqrt{f(\mathrm{Hz})}$.  We note, however, that the observed
dependence of $T_2^*$ on $T_{meas}$ (\figref{omegaIR}) is not
consistent with simple $1/f$ noise.

\section{Temperature Dependence}

\begin{figure}
\resizebox{\columnwidth}{!}{\includegraphics{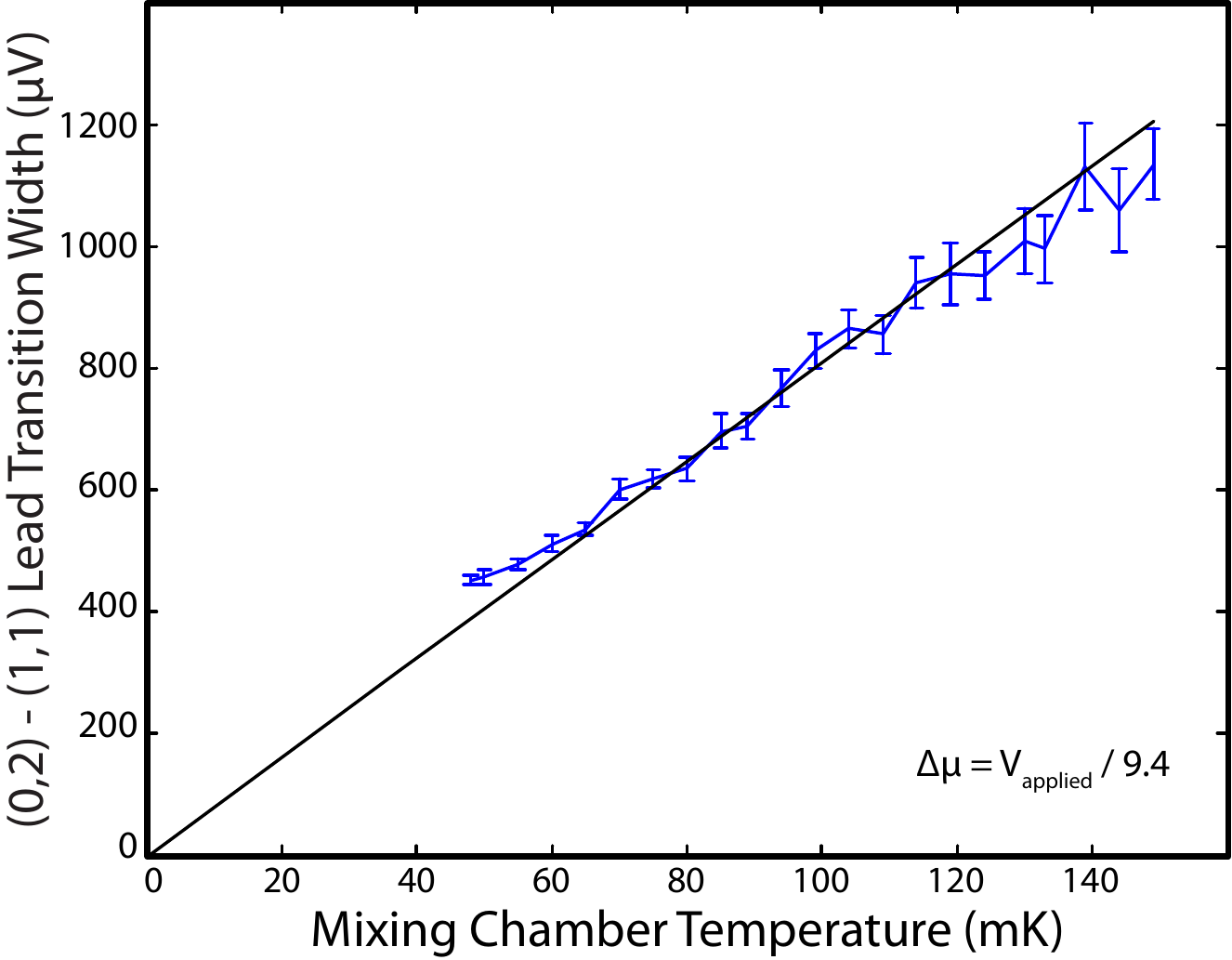}}
\caption{The $(0,1)$-$(0,2)$ charge transition is thermally broadened,
  allowing us to determine the electron temperature.  The width is
  calibrated using the mixing chamber temperature at high
  temperatures.  The black line is a best fit, forced through zero
  width at zero temperature.  The ``lever-arm'' for the fast-RF gates
  is seen to be 9.4, allowing conversion from gate voltage noise to
  effective noise at the qubit if desired.
\label{broadening}
}
\end{figure}

All experiments presented are performed in a dilution refrigerator
with a base temperature of approximately 50mK. All temperature
dependence data are taken by setting the mixing chamber temperature of
the dilution refrigerator and waiting sufficient time (30 min) for the
system to equilbrate. We monitor the electron temperature as the
mixing chamber temperature is changed by measuring the width of a
charge transition of the double QD and assuming that the electronic
temperature and mixing chamber temperatures are the same at high
temperatures (\figref{broadening}).  We find the electronic
temperature and lattice temperature (assumed to be the same as that of
the mixture) are similar over the range studied and we are therefor
unable to differentiate dephasing due to phonons and voltage noise.

\begin{figure}
\resizebox{\columnwidth}{!}{\includegraphics{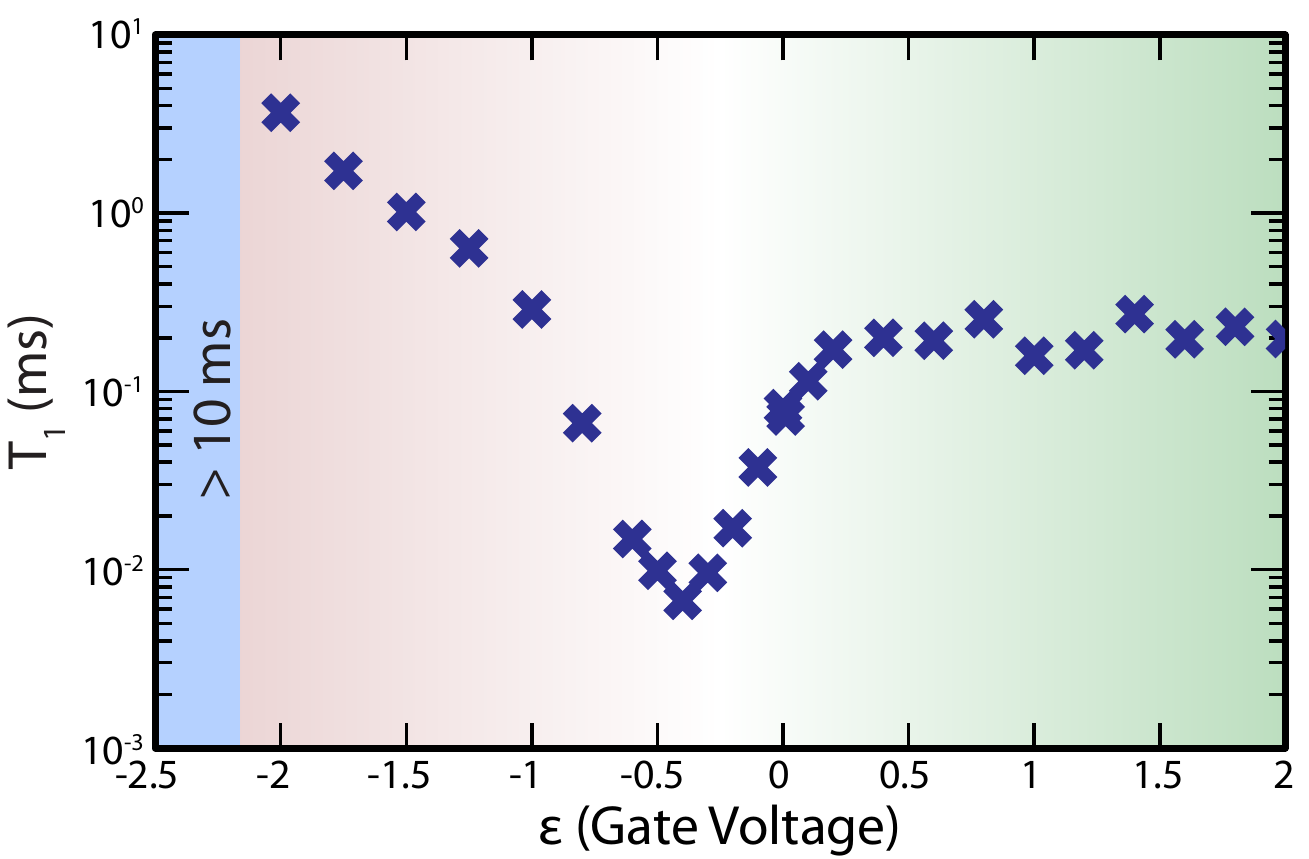}}
\caption{Although $T_1$ is a complex function of $\epsilon$, it is much longer
than $T_2^*$ and $T_2$ everywhere.
\label{t1}
}
\end{figure}
\section{Measuring $T_1$ at finite $\epsilon$}
We measure $T_1$ as a function of $\epsilon$ in two steps.  First we
prepare a \ket{S}, adiabatically ramp to $\epsilon$ and wait for a
time $t$, then measure the probability of the resulting state being a
singlet, $P_{S|S}$.  We then prepare a \ket{T_0} and repeat the
process to measure $P_{S|T_0}$.  We finally fit $P_{S|S}-P_{S|T_0}$ to
an exponential function of $t$ to extract $T_1$.  As seen in
\figref{t1}, $T_1$ is much longer than $T_2$.  The nuclear gradient $\Delta B_Z$ is stabilized to ~30 MHz for this measurement.